%% file: main.tex
\documentclass[twocolumn]{autart}
\usepackage[utf8]{inputenc}

\usepackage{optidef}

\usepackage{amssymb,amsmath,color}
\usepackage{graphicx}    
\usepackage{amsmath, amsfonts}
\usepackage{mathtools}
\usepackage{accents}
\usepackage{url}
\usepackage{tikz,pgfplots}
\usepackage{amsmath}
\usetikzlibrary{arrows}

\usepackage{hyperref}
\usepackage{etoolbox}

\usepackage{xspace,stackrel,hyperref}
\usepackage{mathtools}
\usepackage{tikz,pgfplots}
\usepackage{etoolbox}
\usepackage{autonum}
\usepackage{dsfont}
\usepackage{etoolbox}

\usepackage{multirow}
\usepackage{cite}

\usepackage{arydshln}

\usepackage{esvect}
\usepackage{optidef}
\usepackage{multirow}

\usepackage{algorithm, algpseudocode}
\usepackage{tikzpeople}
\usetikzlibrary{shapes}

\usepackage[textwidth=1.5in]{todonotes}
\DeclareMathOperator*{\argmin}{arg\,min}

\DeclarePairedDelimiter{\ceil}{\lceil}{\rceil}

\newtheorem{assumption}{Assumption}

\newtheorem{theorem}{Theorem}
\newtheorem{proposition}{Proposition}

\newtheorem{definition}{Definition}
\newtheorem{lemma}{Lemma}

\newtheorem{standing assumption}{Standing Assumption}

\pgfplotsset{compat=1.17} 
\usetikzlibrary{pgfplots.groupplots}

\DeclareMathAlphabet{\mathcal}{OMS}{cmsy}{m}{n}

\newcommand{\mc}{\mathcal}

\newcommand{\Z}{\mathbb{Z}}

\newcommand{\R}{\mathds{R}}
\newcommand{\N}{\mathbb{N}}

\definecolor{mycolor1}{RGB}{230,97,1}
\definecolor{mycolor2}{RGB}{178,171,210}
\definecolor{mycolor3}{RGB}{253,184,99}
\definecolor{mycolor4}{RGB}{94,60,153}
\definecolor{mycolor5}{rgb}{0,0,0}

\newcommand{\defineas}{\coloneqq}

\newcommand{\norm}[1]{\left\lVert#1\right\rVert}

\mathtoolsset{showonlyrefs}

\makeatletter 
\pretocmd\@bibitem{\color{black}\csname keycolor#1\endcsname}{}{\fail}
\newcommand\citecolor[1]{\@namedef{keycolor#1}{\color{black}}}
\makeatother 

\begin{document}
\begin{frontmatter}
\title{On decentralized computation of the leader's strategy in bi-level games}
\vspace{-.5cm}
\author{Marko Maljkovic}, \author{Gustav Nilsson}, and \author{Nikolas Geroliminis}

\thanks{The authors are with the School of Architecture, Civil and Environmental Engineering, École Polytechnique Fédérale de Lausanne (EPFL), 1015 Lausanne, Switzerland. {\tt \{marko.maljkovic, gustav.nilsson, nikolas.geroliminis\}@epfl.ch}.}
\thanks{This work was supported by the Swiss National Science Foundation under NCCR Automation, grant agreement 51NF40\_180545.}
\thanks{Some preliminary results of this work were presented in~\cite{ecc2023}.}

\begin{abstract}
\textcolor{black}{
 Motivated by the omnipresence of hierarchical structures in many real-world applications, this study delves into the intricate realm of bi-level games, with a specific focus on exploring local Stackelberg equilibria as a solution concept. While existing literature offers various methods tailored to specific game structures featuring one leader and multiple followers, a comprehensive framework providing formal convergence guarantees to a local Stackelberg equilibrium appears to be lacking. Drawing inspiration from sensitivity results for nonlinear programs and guided by the imperative to maintain scalability and preserve agent privacy, we propose a decentralized approach based on the projected gradient descent with the Armijo stepsize rule. The main challenge here lies in assuring the existence and well-posedness of Jacobians that describe the leader's decision's influence on the achieved equilibrium of the followers. By meticulous tracking of the Implicit Function Theorem requirements at each iteration, we establish formal convergence guarantees to a local Stackelberg equilibrium for a broad class of bi-level games. Building on our prior work on quadratic aggregative Stackelberg games, we also introduce a decentralized warm-start procedure based on the consensus alternating direction method of multipliers addressing the previously reported initialization issues. Finally, we provide empirical validation through two case studies in smart mobility, showcasing the effectiveness of our general method in handling general convex constraints, and the effectiveness of its extension in tackling initialization issues.   
}
\end{abstract} 

\end{frontmatter}
\maketitle
\section{Introduction}
In the realm of strategic decision-making, games with the inherent leader-follower structure have emerged as one of the fundamental frameworks to model the interplay between agents on multiple levels of hierarchy. These games are characterized by a structure in which a leader, possessing a strategic advantage, makes decisions prior to rational followers who, in return, choose their best response to the leader's action. With the pivotal works on bi-level games formalizing the concepts of Stackelberg (SG)~\cite{von1952theory} and their broader format, Reverse Stackelberg games (RSG)~\cite{1102652,HO1982167}, various real-world problems in the domain of energy management~\cite{ex1,AUSSEL2020299}, operational optimization~\cite{ex2,ex3} and transportation~\cite{Groot2017HierarchicalGT, Hierarchical} gained interest from the perspective of computing a no-regret solution for all participants. 

Typically, each lower-level agent competes to minimize a personal objective parametrized by the leader's decision variable and influenced by other followers' decisions. Consequently, the leader aims to minimize a personal objective under the equilibrium constraints imposed by the lower-level game between the followers. If the nature of the application allows for a centralized computation of the solution~\cite{9993196, 10.2307/25614751, 918286}, the problem can be framed as a bi-level mathematical program with complementarity constraints (MPCC)~\cite{10.2307/3690420}, usually tackled by iterative relaxations of the equilibrium constraints~\cite{relax} or by recasting it into an instance of a mixed integer program~\cite{KLEINERT2021100007,bigm}. Nevertheless, in the presence of private feasibility constraints, and driven by the essential requirements to preserve privacy and ensure scalability, decentralized systems have become increasingly prevalent in many real-world applications. As a result, several approaches exploiting specific structural assumptions of the analyzed games have been proposed in the literature. 

\begin{figure}[t]
    \centering
    \resizebox{0.47\textwidth}{!}{
    \begin{tikzpicture}[scale=0.9]
        \shadedraw[top color= mycolor1, bottom color=white, draw=mycolor1] (-1.25, -0.5) rectangle (1.25,0.5);
        \node[scale=0.5] (fl) at (-0,-0){$\begin{array}{rc}
           \text{Decision:}  &  \pi\\
           \text{Objective:}  & J_L\\
           \text{Constraints:} & \mc P
        \end{array}$};
        \node[mycolor1, scale=0.8] (fln) at (0.2, 0.8){Leader};
        \node[businessman, shirt=mycolor1, scale=0.8](b1) at (-0.6, 0.8){};
        
        \draw[] (-1.25, -3.0) rectangle (1.25,-2.0);
        \node[scale=0.7, align=center] (cag) at (0,-2.5) { Communication \\  hub};

        \shadedraw[top color= mycolor2, bottom color=white, draw=mycolor2] (-2.5, -5) rectangle (0.0, -6);
        \node[scale=0.5] (f2p) at (-1.25,-5.5){$\begin{array}{rc}
           \text{Decision:}  &  x_2\\
           \text{Objective:}  & J_2\\
           \text{Constraints:} & \mc X_2(x_2,\pi_t)
        \end{array}$};
        \node[mycolor2, scale=0.7] (f2n) at (-0.95, -6.25){Follower 2};
        \node[businessman, shirt=mycolor2, scale=0.7](b2f) at (-1.75, -6.25){};
        
        \shadedraw[top color= mycolor2, bottom color=white, draw=mycolor2] (-5.1, -5) rectangle (-2.6, -6);
        \node[scale=0.5] (f1p) at (-3.85,-5.5){$\begin{array}{rc}
           \text{Decision:}  &  x_1\\
           \text{Objective:}  & J_1\\
           \text{Constraints:} & \mc X_1(x_1,\pi_t)
        \end{array}$};
       \node[mycolor2, scale=0.7] (f1n) at (-3.55, -6.25){Follower 1};
        \node[businessman, shirt=mycolor2, scale=0.7](b1f) at (-4.35, -6.25){};

        \shadedraw[top color= mycolor2, bottom color=white, draw=mycolor2] (2.6, -5) rectangle (5.1, -6);
        \node[scale=0.5] (fnp) at (3.85,-5.5){$\begin{array}{rc}
           \text{Decision:}  &  x_N\\
           \text{Objective:}  & J_N\\
           \text{Constraints:} & \mc X_N(x_N,\pi_t)
        \end{array}$};
        \node[mycolor2, scale=0.7] (f1n) at (4.15, -6.25){Follower $N$};
        \node[businessman, shirt=mycolor2, scale=0.7](b1f) at (3.30, -6.25){};
        
        \node (d1) at (1.3, -5.5)[circle,fill,inner sep=0.75pt]{};
        \node (d2) at (1.5, -5.5)[circle,fill,inner sep=0.75pt]{};  \node (d3) at (1.1, -5.5)[circle,fill,inner sep=0.75pt]{};   
        
        \draw[dashed] (-5.2, -4.5) rectangle (5.2, -6.5);

        \draw[->] (-3.85 , -5) -- (-0.75,-3);
        \draw[->] (-1.25 , -5) -- (-0.25,-3);
        \draw[->] (3.85  , -5) -- (0.75,-3);
        
        \draw[->] (0.0, -3) -- (0.0, -4.5);
        
        \node[scale=0.8] (D1) at (-3.0, -3.85){$\textbf{D}_{\pi_t}x^{*}_1$};
        \node[scale=0.8] (D2) at (-1.3, -3.85){$\textbf{D}_{\pi_t}x^{*}_2$};
        \node[scale=0.8] (D3) at (0.3, -3.85){$\pi_t$};
        \node[scale=0.8] (D4) at (3.0, -3.85){$\textbf{D}_{\pi_t}x^{*}_N$}; 
        
        \draw[-] (1.25, -2.5) -- (1.75, -2.5);
        \draw[-] (1.75, -2.5) -- (1.75, 0.0);
        \draw[->] (1.75, 0.0) -- (1.25,0.0);
        
        \draw[-] (-1.25,0.0) -- (-1.75, 0.0);
        \draw[-] (-1.75, 0.0) -- (-1.75, -2.5);
        \draw[->] (-1.75, -2.5) -- (-1.25, -2.5);
        
        \node[scale=0.8] (Dt) at (2.35, -1.25){$\textbf{D}_{\pi_t}x^{*}$};       
        \node[scale=0.8] (pi) at (-2.15, -1.25){$\pi_{t+1}$};
        
        
    \end{tikzpicture}}

    \caption {Schematic sketch of the problem setup. Each of the $N$ followers aims to optimize the personal objective $J_i$ under the parametrized local constraints $x_i\in\mc X_i(x_i,\pi)$. The followers communicate with the leader through the communication hub that is used as a medium to collect the locally computed Jacobians $\textbf{D}_{\pi_t}x^{*}_i$ in every update step of the leader's action.}
    \label{fig:problem}
\end{figure}
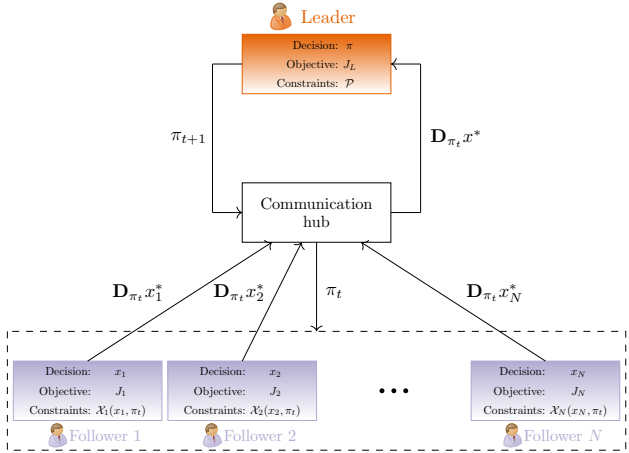

For a specific class of pricing games with a quadratic, aggregative game between the followers~\cite{ecc2022, Hierarchical}, it has been demonstrated how the concept of Reverse Stackelberg games can be used to incentivize the global optimum of the leader. Formulating dynamic strategies for the leader in the form of functionals, rather than real-valued vectors, facilitated reshaping the lower-level game in a way that gave rise to a Nash Equilibrium corresponding exactly to the minimizer of the leader's objective. A follow-up question naturally imposes - can we establish a connection between the proposed dynamic policies and fixed, real-valued, vector strategies? In general, this procedure is not straightforward. By restricting the leader's impact solely to parameterizing the lower-level game, rather than allowing the flexibility to restructure it as in the RSG framework, we enter the realm of Stackelberg games, which are arguably more challenging to solve. 

Due to the inherent non-convex nature of the bi-level problems, the existing body of literature predominantly focuses on local Stackelberg equilibria as a viable solution concept~\cite{9424958}. In~\cite{7956147}, an iterative method for unconstrained games has been suggested. Conversely, in the context of constrained, quadratic, aggregative games, the authors of~\cite{9424958} propose a two-layer, semi-decentralized algorithm based on iteratively convexifying a regularized version of the underlying MPCC. Looking from a different perspective, the sensitivity results for nonlinear programs~\cite{pmlr-v48-pedregosa16,9468930,Wang_Xu_Perrault_Reiter_Tambe_2022} hint at the possibility of differentiating the Karush-Kuhn-Tucker (KKT) conditions of the best-response optimization problems in an attempt to estimate how the attained Nash Equilibrium between the followers reacts to a change in the leader’s action. In light of the success that gradient-descent-based methods have experienced in many real-world applications and under the paradigm of decentralized equilibrium computation, we aim to design an iterative, first-order-like method for computing a local Stackelberg equilibrium suitable for a broad class of bi-level games.  

This paper is a continuation of the preliminary work presented in~\cite{ecc2023}, where the original idea inspired by~\cite{Wang_Xu_Perrault_Reiter_Tambe_2022} has been outlined for one specific case of quadratic, aggregative Stackelberg games. In this paper, we extend the analysis to 
a more general structure illustrated in Figure~\ref{fig:problem}, and show how the computation of Jacobians describing the influence of the leader's strategy on the attained variational Nash Equilibrium~\cite{ecc2023} of the lower-level game can be performed locally by each of the followers. We start by briefly discussing the connection between the RSG dynamic policies and the static ones used in the SG setup and continue by rigorously tackling the requirements of the Implicit Function Theorem~\cite{Implicit} in order to generalize the approach in~\cite{ecc2023} to also account for non-quadratic, non-aggregative games. On the other hand, in the context of quadratic, aggregative games with polytopic constraints as in~\cite{ecc2023}, we also address the reported initialization issue arising from the fact that in each iteration we differentiate the KKT conditions of an optimization problem equivalent to the standard best-response one. With that in mind, the main contributions of this paper can be summarized as follows:
\begin{itemize}
    \item We propose a distributed, first-order-like, iterative method based on explicit fulfillment of the Implicit Function Theorem requirements. By ensuring a local improvement of the leader’s objective at each iteration, we provide formal convergence guarantees for a broad class of bi-level games.  
    \item For a class of quadratic, aggregative, Stackelberg games linearly parametrized by the decision variable of the leader, we propose a decentralized warm-start procedure based on the alternating direction method of multipliers (ADMM). In line with the distributed nature of the main algorithm, we compute a feasible leader's strategy that yields an interior point variational Nash Equilibrium of the lower-level game, i.e., renders no local inequality constraint active.
    \item We test the proposed method in an adaptation of the case study from the domain of smart mobility previously analyzed in~\cite{ecc2023}. Firstly, by introducing the so-called budget constraints parametrized by the leader's decision variable, we demonstrate the effectiveness of the main procedure for the bi-level games with general convex constraints. Then, by going back to the setup in~\cite{ecc2023}, we illustrate the effectiveness of the warm-start procedure in alleviating the initialization issues.  
\end{itemize}

The paper is outlined as follows: the rest of this section
is devoted to introducing some basic notation. In Section~\ref{sec:tp}, we introduce the general bi-level setup, discuss the connection between the Stackelberg and Reverse Stackelberg games, postulate the main standing assumptions, and formally introduce the problem. In the following section, Section~\ref{sec:3}, we revise and generalize the decentralized method for computing the local Stackelberg equilibrium previously outlined in~\cite{ecc2023}. Section~\ref{sec:LQG} then focuses on the subclass of quadratic, aggregative Stackelberg games and presents the proposed warm-start procedure. Finally, we conclude the paper with Sections~\ref{sec:5} and~\ref{sec:6} where we present the numerical examples and propose some ideas for future research.

\textit{Notation}: Let $\R$ denote the set of real numbers, $\R_+$ the set of non-negative reals, and $\Z_+$ the set of non-negative integers. Let $\mathbf{0}_{m}$ and $\mathbf{1}_{m}$ denote the all zero and all one vectors of length $m$ respectively, and $\mathbb{I}_{m}$ the identity matrix of size $m \times m$. For a finite set $\mc A$, we let $\R_{(+)}^{\mc A}$ denote the set of (non-negative) real vectors indexed by the elements of $\mc A$ and $\left|\mc A\right|$ the cardinality of $\mc A$. Furthermore, for finite sets $\mc A$, $\mc B$ and a set of $|\mc B|$ vectors $x_i\in \R_{(+)}^{\mc A}$, we define $x \defineas \text{col}((x_{i})_{i\in \mc B})\in \R^{|\mc A||\mc B|}$ to be their concatenation. For $A\in \R^{n\times n}$, $A\succ 0 (\succeq 0)$ is equivalent to $x^TAx>0 (\geq 0)$ for all $x\in\R^{n\times n}$. We let $A\otimes B$ denote the Kronecker product between two matrices and for a vector $x\in\R^n$, we let $\text{Dg}(x)\in\R^{n\times n}$ denote a diagonal matrix whose elements on the diagonal correspond to vector $x$. For a differentiable function $f(x):\R^n\rightarrow \R^m$, we let $\textbf{D}_xf\in\R^{m\times n}$ denote the Jacobian matrix of $f$ defined as $(\textbf{D}_x f)_{ij}\defineas\frac{\partial f_i}{\partial x_j}$. If $f(x)$ is a real-valued function, i.e., $m=1$, we adopt $\nabla_x f\defineas\textbf{D}_xf\in\R^n$. Finally, for a set-valued mapping $\mathcal{F}: \mc Y \rightrightarrows \mc X, \operatorname{gph}(\mathcal{F}):=$ $\left\{(y, x) \in \mc Y \times \mc X \mid x \in \mathcal{F}(y)\right\}$ denotes its graph.

\section{Theoretical preliminaries}\label{sec:tp}
Throughout the paper, we consider a bi-level game with a set of $N+1$ agents $\overline{\mc I}=\mc I\cup\left\{L\right\}$, where $L$ represents the leading agent and each $i\in\mc I$ represents one of the $N$ followers. In this setup, the leading agent is the first one to choose an action $\pi$ from its feasible set $\mc P$, to which all $N$ followers will respond at once with a personal decision vector $x_i$ from their feasible set $\mc X_i(\pi)$ that is in their best interest. If $m_F\in\N$ represents the dimension of the follower's decision space, we assume $\mc X_i(\pi)$ in the form of
$\mc X_i(\pi)\defineas\{x_i\in\R^{m_F}\mid \boldsymbol{g}^{\operatorname{inq}}_i(x_i,\pi)\leq\mathbf{0}_{m_{\operatorname{inq},i}}\land\boldsymbol{g}^{\operatorname{eq}}_i(x_i,\pi)=\mathbf{0}_{m_{\operatorname{eq},i}}\}$, where $m_{\operatorname{inq},i},m_{\operatorname{eq},i}\in\N$ denote the number of inequality and equality constraints encompassed in $\mc X_i(\pi)$. If $m_L\in\N$ represents the dimension of the leader's action, the nature of the leader's strategy can lead to two types of games in general:
\begin{itemize}
    \item \textbf{Reverse Stackelberg Games (RSG)}: where the leader's strategy $\pi\in\mc P$ is a map $\pi: \R^{Nm_F}\rightarrow \R^{m_L}$;
    \item \textbf{Stackelberg Games (SG)}: where the leader's strategy is a fixed real vector, $\pi\in\mc P\subseteq \R^{m_L}$.
\end{itemize}
In any case, we refer to the phase of choosing the optimal $x_i\in\mc X_i(\pi)$ as the \textit{Lower-level game} and the process of choosing the optimal leader's strategy knowing that the followers will play a best-response as the \textit{Upper-level game}. Furthermore, we define the joint strategy of all followers as $x \defineas \text{col}((x_{i})_{i\in \mc I})\in\mc X(\pi)$ and for every $i\in\mc I$, we define $x_{-i} \defineas \text{col}((x_{j})_{j\in \mc I \setminus i})\in\mc X_{-i}(\pi)$, such that $\mc X(\pi)\defineas \prod_{i\in\mc I} \mc X_{i}(\pi)$ and $\mc X_{-i}(\pi)\defineas \prod_{j\in\mc I\setminus i} \mc X_{j}(\pi)$. 
\subsection{Lower-level game}\label{subsec:llg}
Regardless of the game type, the followers choose their strategies in an attempt to minimize personal objective functions $J_i(x_i, x_{-i}, \pi)$ by playing the best response to other followers' strategies under the umbrella of the Nash Equilibrium concept. The $\pi$-parametrized lower-level game $G^0(\mc I;\pi)$ represents $N$ coupled optimization problems, i.e., $G^0(\mc I;\pi)\defineas\{G^0_i(\pi,x_{-i})\:\:|\:\:i\in\mc I\}$, with
\begin{equation}
    G^0_{i}\left(\pi,x_{-i}\right):=\left\{\begin{array}{c}
    \displaystyle\min_{x_i\in\R^{m_F}}J_i\left(x_i,x_{-i},\pi\right) \\
    \text { s.t. }x_i\in\mc X_i(\pi)
    \end{array}\right\} \,,
    \label{eq:best_resp}
\end{equation}
and the Nash Equilibrium given in Definition~\ref{def:1}.
\begin{definition}[Nash Equilibrium]\label{def:1}
    For any leader's strategy $\pi\in\mc P$, a joint strategy $x^*\in\mc X$ is a Nash Equilibrium (NE) of the game $G^0(\mc I;\pi)$, if for all $i\in\mc I$ and all $x_i\in\mc X_i(\pi)$ it holds that $J_i\left(x_{i}^*, x_{-i}^*, \pi\right)\leq J_i\left(x_{i}, x_{-i}^*,\pi\right)$.
\end{definition}
For a particular $\pi\in\mc P$, it is rarely possible to find a closed-form characterization of the full set of NE in a general setup. Therefore, we postulate standard assumptions about the structure of $G^0(\mc I;\pi)$ that allow us to focus on the variational Nash Equilibria (v-NE) as a solution concept of the lower-level game. 
\begin{standing assumption}\label{sass:1}
For every $i\in\mc I$ and any $\pi\in\mc P$, $x_{-i}\in\mc X_{-i}$, the cost $J_i(x_i,x_{-i},\pi)$ is convex and continuously differentiable in $x_i$. Moreover, it is continuous in $x\in\mc X(\pi)$ and the sets $\mc X_i(\pi)$ are nonempty, compact, convex and satisfy Slater's constraint qualification. 
\end{standing assumption}
Strictly speaking, under Standing Assumption~\ref{sass:1}, for every $i\in\mc I$, the Nash Equilibrium strategy $x^{*}_i\in\mc X_i(\pi)$ is the solution of the best-response optimization problem~\eqref{eq:best_resp} for $x^{*}_{-i}$, i.e., $G^0_{i}(\pi,x_{-i}^*)$. The optimality of $x^{*}_i\in\mc X_i(\pi)$ is guaranteed if and only if $x^{*}_i$ solves the KKT system of equations $l_i(z_i,\pi\mid x_{-i}^*)=0$, where the vector mapping $l_i$ is defined for every $z_i=(x_i,\lambda_i,\nu_i)$ as
\begin{equation}
\label{eq:KKToperator}
    l_i\left(z_i,\pi\mid x_{-i}^*\right)\defineas\left[\begin{array}{c}\nabla_{x_i}\mc L_i\left(z_i,\pi\right) \\ \operatorname{Dg}\left(\lambda_i\right)\boldsymbol{g}_i^{\operatorname{inq}}\left(x_i,\pi\right) \\ \boldsymbol{g}_i^{\operatorname{eq}}\left(x_i,\pi\right)\end{array}\right]\,,
\end{equation}
with the Lagrangian given by $\mc L_i(z_i,\pi)=J_i(x_{i}, x_{-i}^*, \pi)+\lambda_i^T\boldsymbol{g}_i^{\text{inq}}(x_{i},\pi)+\nu_i^T\boldsymbol{g}_i^{\operatorname{eq}}(x_{i},\pi)$
and $\lambda_i\in\R_+^{m_{\operatorname{inq},i}}$ and $\nu_i\in\R^{m_{\operatorname{eq},i}}$ representing the dual variables associated with the inequality and equality constraints. Under Standing Assumption~\ref{sass:1}, if some $\hat{z}_i=(\hat{x}_i,\hat{\lambda}_i,\hat{\nu}_i)$, with feasible $\hat{x}_i$ and $\hat{\lambda}_i$, satisfies $l_i\left(\hat{z}_i,\pi\mid x_{-i}^*\right)=0$, then $\hat{z}_i$ is the optimizer of~\eqref{eq:best_resp}. On the other hand, based on~\cite[Prop. 1.4.2]{VIProblems}, Standing Assumption~\ref{sass:1} also ensures that a joint strategy $x\in\mc X(\pi)$ is a NE if and only if it solves a variational inequality problem, hence providing a closed-form description of the lower-level game's solution set.  Namely, if
$F(x,\pi)\defineas\text{col}((\nabla_{x_{i}}J_{i}(x_i, x_{-i},\pi))_{i\in \mc I})$ denotes the pseudo-gradient of $G^0(\mc I;\pi)$, we can adopt the following assumption about the followers.
\begin{standing assumption}\label{sass:2}
For any $\pi\in\mc P$, the agents $i\in\mc I$ play a joint strategy $x\in\mc N_0(\pi)$, where $\mc N_0(\pi)$ is the set of all v-NE of the game $G^0(\mc I;\pi)$, given by $ \mc N_0(\pi)\defineas\{x\in\mc X(\pi)\mid(y-x)^TF(x,\pi)\geq 0,\:\forall y\in\mc X(\pi)\}$.
\end{standing assumption}
\subsection{Upper-level game}\label{subsec:blg}
On the upper level, finding the optimal strategy $\pi\in\mc P$ imposes solving a minimization problem of the leader's objective $J_L:\mc \R^{m_F}\times\mc P\rightarrow\R$. Instances of both SG and RSG can be compactly written as:
    \begin{equation}
        G_1:=\left\{\begin{array}{c}
        \displaystyle\min _{\pi \in\mc P}  J_{L}\left(x^{*},\pi\right) \\
        \text { s.t. } \left(x^{*},\pi\right) \in\text{gph}\left(\mc N_0\right)\cap\left(\R^{m_F}\times\mc P\right)
        \end{array}\right\} \,.
        \label{eq:SG}
    \end{equation}
In general, the optimal $\pi$ in $G_1$ is a possibly non-unique solution~\cite[Corr. 1]{ecc2023} to a non-convex problem that requires the ability to understand the leader's influence on the position of the lower-level game's v-NE.  Moreover, depending on the properties of $F(x,\pi)$, the lower-level game could admit multiple NE for a particular parametrization. In this study, we restrict ourselves to cases where $G^0(\mc I;\pi)$ admits a unique NE for any $\pi\in\mc P$. Hence, we state the following assumption, common in existing literature~\cite{BIANCHI2022110080,Paccagnan2016b}, that ensures the existence and uniqueness of the lower-level game's v-NE~\cite[Th. 2.3.3]{VIProblems}. Moreover, for a particular leader's strategy, it also ensures that the v-NE can be computed as a fixed-point of the projected pseudo-gradient mapping~\cite{VIProblems, Hierarchical, Paccagnan2016a, Paccagnan2019}.
\begin{standing assumption}\label{sass:3}
For any $\pi\in\mc P$, the pseudo-gradient $F(\cdot,\pi)$ is strongly monotone in $x\in\mc X(\pi)$. 
\end{standing assumption}
Concerning the nature of the leader's decision variable, it is evident that SG represents a distinct instance of RSG, wherein the leader's strategy assumes a constant function. This limits the flexibility to incentivize a certain NE of the lower-level game, as the leader's strategies in the form of feedback policies offer a means to directly shape the functional form of the followers' optimization problems. Moreover, with many real-world applications requiring different notions of fairness, solving a SG can be considered arguably more challenging. To elucidate this contrast, we look at the following example for a specific class of games, referred to as quadratic aggregative games in~\cite{Paccagnan2016a,Hierarchical}.  
\begin{definition}[Quadratic Aggregative Games]\label{ex:1}
Let the leaader's objective be $J_L=\frac{1}{2}\sigma(x)^{T}P_L\sigma(x)+q_L^T\sigma(x)$, where $\sigma(x)=\sum_{i\in\mc I}x_i$. Moreover, let the lower-level game $G^0(\mc I;\pi)$ be defined by 
\begin{equation}\label{eq:ji}
    J_i(x_i,x_{-i},\pi)=\frac{1}{2}x_i^TP_ix_i+x_i^TQ_i\sigma(x_{-i})+r_i^Tx_i+x_i^TS_i\pi_i\,,
\end{equation}
where $\sigma(x_{-i})=\sigma(x)-x_i$, $\pi_i\in\R^{m_F}$, $\pi=\text{col}((\pi_i)_{i\in\mc I})$, the matrices $P_L$, $P_i$, $Q_i$, $S_i$ and vectors $q_L$, $r_i$ are all real valued, $P_L\succ0$, for every $i\in\mc I$, $P_i,S_i\succ0$, and the Standing assumptions~\ref{sass:1},~\ref{sass:2} and~\ref{sass:3} all hold.
\end{definition}
If one regards the game in Definition~\ref{ex:1} as a Reverse Stackelberg game, it suffices to choose the leader's strategy as a mapping of the form
\begin{equation}
    \pi_i(x_{i},x_{-i})=S_i^{-1}\left[\frac{1}{2}\overline{\textbf{P}}_ix_i+\overline{\textbf{Q}}_i\sigma(x_{-i})+\overline{\textbf{r}}_i\right]\,,
    \label{eq:rsgp}
\end{equation}where $\overline{\textbf{P}}_i=P_L-P_i$, $\overline{\textbf{Q}}_i=P_L-Q_i$ and $\overline{\textbf{r}}_i=q_L-r_i$, so that the leader's objective and the objectives of the followers satisfy
$J_L(x_i,x_{-i})-J_L(\tilde{x}_i,x_{-i})=J_i(x_i,x_{-i})-J_i(\tilde{x}_i,x_{-i})$ for any fixed $x_{-i}\in\mc X_{-i}$ and any two $x_i, \tilde{x}_i\in\mc X_i$. This implies that due to~\eqref{eq:rsgp}, the leader's objective $J_L$ becomes the exact potential~\cite{PotentialGames} of the lower-level game by definition. Consequently, this guarantees that the minimizer of $J_L$ aligns with the v-NE of $G^0(\mc I;\pi)$, which can, in this case, be computed using a decentralized, iterative, fixed-point method as in~\cite{ecc2023,Hierarchical,Paccagnan2016a}.
Conversely, if one regards the game in Definition~\ref{ex:1} as an instance of Stackelberg games, such manipulation is no longer possible. If $x^{R}$ represents the NE obtained when applying~\eqref{eq:rsgp} in the RSG setup, one might naively try to plug back $x^{R}$ into~\eqref{eq:rsgp} to obtain a static pricing vector $\pi_i^{R}=\pi_i(x^{R}_i,x^{R}_{-i})$ and use it in the setup of a SG. However, in a general case, this does not yield a viable solution, as illustrated in the following proposition.
\begin{proposition}\label{prop:1}
Let a bi-level game be defined as in Definition~\ref{ex:1} such that $\mc X_i(\pi)=\R^{m_F}$ for all $i\in\mc I$. Moreover, let the mapping $\pi:\R^{Nm_F}\rightarrow\R^{Nm_F}$ be given by $\pi_i:\R^{Nm_F}\rightarrow\R^{m_F}$ and let~\eqref{eq:rsgp} yield $x^R\in\mc X(\pi)$. If $\pi_i^{R}=\pi_i(x^{R}_i,x^{R}_{-i})$, then utilizing $\pi_i^{R}$ in a SG gives rise to a NE equal to $x^R$ if and only if $P_Lx_i^R=P_ix_i^R$.
\end{proposition}
\begin{pf}\label{ap:cor1} 
It suffices to look at the KKT systems $l_i(z_i,\pi_i\mid x_{-i}^{R})=0$, given by~\eqref{eq:KKToperator}, for the SG and RSG scenarios. Namely, after applying~\eqref{eq:rsgp}, the derivative of the $i$-th follower's Lagrangian evaluated at $x^R_i$ satisfies  $P_Lx^R_i+P_L\sigma(x_{-i}^R)+q_L=0$. On the other hand, if $\pi^R_i$ is applied in the context of SG, $x^R_i$ will remain the NE of $G^0(\mc I;\pi)$ if and only if $\frac{1}{2}(P_L+P_i)x^R_i+P_L\sigma(x_{-i}^R)+q_L=0$ holds. Hence, $x^R$ remains the NE of $G^0(\mc I;\pi)$ if and only if $\frac{1}{2}P_Lx_i^R=\frac{1}{2}P_ix_i^R$, which does not always hold. \qed
\end{pf}
Therefore, even the Stackelberg games that exhibit favorable mathematical properties such as the one in Definition~\ref{ex:1} pose significant difficulties in computing the leader's strategy. As previously mentioned in the introduction, if the nature of the application allows centralized computation, one can formulate an MPCC that can be recast into an instance of mixed-integer linear or quadratic problems~\cite{KLEINERT2021100007} using the big-M reformulation~\cite{bigm} as demonstrated in~\cite{918286, 9993196, maljkovic2023learning}. Unfortunately, such computation could breach the privacy of the lower-level agents in many real-world applications, particularly in terms of sharing information about personal constraint sets $\mc X_i$. With that in mind, the focus of this paper is entirely redirected towards the decentralized computation of the leader's strategy in Stackelberg games.
\subsection{Problem formulation}\label{sec:pf}
Owing to the problem's overall non-convex character, with possibly multiple solutions, in this work, we focus on finding the leader's strategy $\pi\in\mc P\subseteq\R^{m_F}$ based on the concept of local Stackelbrg equilibria (l-SE) previously explored in~\cite{7088583,10.2307/25147124,9424958}.
For the sake of completeness, we repeat it in Definition~\ref{def:lSE}.
\begin{definition}[Local Stackelberg Equilibrium]\label{def:lSE}
    Let $G_1$ be a Stackelberg game as in~\eqref{eq:SG}. A pair of vectors $\left(\hat{x}^*,\hat{\pi}\right)\in\operatorname{gph}\left(\mc N_0\right)\cap\left(\R^{m_F}\times\mc P\right)$ is a local Stackelberg equilibrium of $G_1$ if there exist open neighborhoods $\Omega_{\hat{x}^*}$ and $\Omega_{\hat{\pi}}$ of $\hat{x}^*$ and $\hat{\pi}$ respectively, such that
    \begin{equation}
        \label{eq:lSE}
        J_L\left(\hat{x}^*,\hat{\pi}\right)\leq\inf_{\left(x^*,\pi\right)\in\operatorname{gph}\left(\mc N_0\right)\cap\Omega}J_L\left(x^*,\pi\right)\,,
    \end{equation}
    where $\Omega\defineas\Omega_{\hat{x}^*}\times\left(\mc P\cap\Omega_{\hat{\pi}}\right)$.
\end{definition}
Interestingly, restricting ourselves to the framework of Stackelberg games implies that finding the leader's strategy for the bi-level game~\eqref{eq:SG} in the context of l-SE reduces to finding the local optimum of $J_L$ as a function of $\pi$. Namely, under Standing Assumption~\ref{sass:1}, for any $\pi\in\mc P$, we have that $|\mc N_0(\pi)|=1$. This means that to find the l-SE, we need to find $\hat{\pi}$ and its neighborhood $\Omega_{\hat{\pi}}$, since the condition~\eqref{eq:lSE} will always be fulfilled for $\hat{x}^*=\mc N_0(\hat{\pi})$ and the open ball of radius $R$ given by   $\Omega_{\hat{x}^*}\defineas\left\{x\in\mc X \mid \norm{x-\hat{x}^*}<R\right\}$, where  $R>\max_{x\in\hat{\mc N}\left(\Omega_{\hat{\pi}}\right)}\norm{x-\hat{x}^*}$ and $\hat{\mc N}(\Omega_{\hat{\pi}})=\bigcup_{\pi\in\Omega_{\hat{\pi}}}\mc N_0(\pi)$. 

To summarize, in the following sections we will focus on designing an iterative, decentralized, gradient descent-based algorithm that leverages the guarantees provided by the Implicit Function Theorem~\cite{Implicit} concerning the continuous differentiability of $J_L(x^*(\pi),\pi)$ at the current $\pi$ value. However, before delving deeper into the details, we establish the regularity of the leader's optimization problem through Standing Assumption~\ref{sass:4}.
\begin{standing assumption}\label{sass:4}
The leader's constraint set $\mc P\subseteq\R^{m_L}$ is nonempty, compact and convex. Moreover, $J_L:\R^{Nm_F}\times\mc P\rightarrow\R$ is continuously differentiable in both $x^*$ and $\pi$, and for every $i\in\mc I$, each element of $\boldsymbol{g}^{\operatorname{inq}}_i(x_i,\pi)$ and $\boldsymbol{g}^{\operatorname{eq}}_i(x_i,\pi)$ is continuously differentiable in both $x_i$ and $\pi$. Finally, for every $x_{-i}^*\in\mc X_{-i}(\pi)$, every component of the derivative of the Lagrangian associated with the KKT system $l_i(z_i,\pi\mid x_{-i}^*)=0$, i.e., $\nabla_{x_i}\mc L(z_i,\pi)$, is continuously differentiable at both $x_i$ and $\pi$.
\end{standing assumption}
\section{Decentralized computation of the local Stackelberg equilibrium}\label{sec:3}
To tackle the problem of computing the local Stackelberg equilibrium, the initial step involves introducing the idea of Projected Gradient descent incorporating the Armijo rule. This concept, along with the Implicit Function Theorem, will form the foundation of our method.
\subsection{Projected Gradient descent with Armijo rule}
We start by adopting the projected gradient descent method with 'Armijo step-size rule along the projection arc' explored in~\cite{Bertsekas/99}. To update $\pi$ at iteration $t\in\N$, we first define the mapping $\pi^+:\mc P\times\R_{+}\rightarrow\mc P$ as
\begin{equation}
\label{eq:pgd1}
    \pi^{+}\left(\pi_t, s\right)\defineas \Pi_{\mc P}\left[\pi_t-\left.s\frac{\text{d} J^L\left(x^*_{\pi},\pi\right)}{\text{d}\pi}\right\vert_{\pi=\pi_t}\right]\,,
\end{equation}
where $\Pi_{\mc P}$ is the projection operator on the leader's constraint set for some particular step size $s\in\R_+$ and $x^*_{\pi}$ emphasizes the dependence of the Nash Equilibrium on $\pi$. Let $\beta$, $\overline{s}$ and $\delta$ be fixed scalars such that $\beta,\delta\in\left(0,1\right)$ and $\overline{s} >0$. Moreover, let $l_t\in\Z_{\geq0}$ be the smallest non-negative integer such that for $s_t=\beta^{l_t}\overline{s}$ it holds that
\begin{equation}
\label{eq:pgd3}
    \label{eq:termcond}
    \begin{split}
    &J^L\left(x^*_{\pi_t},\pi_t\right)-J^{L}\left(x^*_{\pi^+\left(\pi_t, s_t\right)},\pi^+\left(\pi_t, s_t\right)\right)\geq \\
    & \geq\delta\left(\left.\frac{\text{d} J^L\left(x^*_{\pi},\pi\right)}{\text{d}\pi}\right\vert_{\pi=\pi_t}\right)^T\left(\pi_t-\pi^{+}\left(\pi_t, s_t\right)\right)\,.
    \end{split}
\end{equation}
Then, the leader's strategy is updated as
\begin{equation}
\label{eq:pgd2}
    \pi_{t+1}=\pi^{+}(\pi_t, s_t)\,.
\end{equation}
Under Standing Assumption~\ref{sass:4}, to observe that $l_t$ is well defined, i.e., a stepsize $s_t$ will be found after a finite number of trials based on the test given by~\eqref{eq:pgd3}, it suffices to invoke the following adaptation of~\cite[Prop. 2.3.3]{Bertsekas/99}.
\begin{lemma}[Proposition 2.3.3 of~\cite{Bertsekas/99}]\label{lem:1}
    Let the set $\mc P$ satisfy Standing Assumption~\ref{sass:4}, $J_L(x^*_{\pi},\pi)$ be continuously differentiable on $\mc P$ and $\delta\in(0,1)$. Then, for every $\pi\in\mc P$, there exists $s_{\pi}>0$ such that $J_L(x^*_{\pi},\pi)-J_L(x^*_{\pi^+(\pi,s)},\pi^+(\pi,s))\geq\delta\nabla_{\pi} J_L(x^*_{\pi},\pi)^T(\pi-\pi^+(\pi,s)))$ holds for every $s\in[0,s_{\pi}]$.
\end{lemma}
Therefore, the complexity of each update step boils down to ensuring that $J_L(x^*_{\pi},\pi)$ is continuously differentiable, i.e., showing that the gradient of the leader's objective with respect to the current strategy given by
\begin{equation}
\label{eq:derivative1}
    \frac{\text{d} J_L\left(x^*_{\pi},\pi\right)}{\text{d} \pi}=\frac{\partial J_L\left(x^*_{\pi},\pi\right)}{\partial \pi}+\textbf{D}^T_{\pi}x^{*}_\pi\frac{\partial J_L\left(x^*_{\pi},\pi\right)}{\partial x^*_\pi}\,,
\end{equation}
is well-defined. In that case, if $\overline{s}>s_{\pi_t}$ we have $l_t=0$, otherwise the testing procedure~\eqref{eq:pgd3} terminates after $l_t=\ceil{\log_{\beta}(\frac{s_{\pi}}{\overline{s}})}$ iterations. 

The challenging aspect of computing~\eqref{eq:derivative1} stems from the requirement to compute the Jacobian $\textbf{D}_{\pi}x^{*}_\pi$, i.e., from having to estimate how the NE of $G^0(\mc I;\pi)$ reacts to variations in $\pi$. This is particularly challenging as in general there exists no closed-form functional description of the connection between $\pi$ and the obtained NE $x^*_\pi$. Therefore, we aim to achieve this by virtue of the Implicit Function Theorem. Namely, with the constraint sets $\mc X_i(\pi)$ being local, and knowing that 
\begin{equation}
    \label{eq:separate}
    \textbf{D}^T_{\pi}x^{*}_\pi\frac{\partial J_L\left(x^*_{\pi},\pi\right)}{\partial x^*_\pi}=\sum_{i\in\mc I}\textbf{D}^T_{\pi}x^{*}_{\pi,i}\frac{\partial J_L\left(x^*_{\pi},\pi\right)}{\partial x^{*}_{\pi,i}}\,,
\end{equation}
we can compute $\textbf{D}_{\pi}x^{*}_\pi$ in a distributed manner such that each follower remains in charge of only computing the personal Jacobian $\textbf{D}_{\pi}x^{*}_{\pi,i}$. The Jacobians are then communicated to the leader as illustrated in Figure~\ref{fig:problem}, who, in return, calculates 
\begin{equation}
\label{eq:derivative2}
    \frac{\text{d} J_L\left(x^*_{\pi},\pi\right)}{\text{d} \pi}=\frac{\partial J_L\left(x^*_{\pi},\pi\right)}{\partial \pi}+\sum_{i\in\mc I}\textbf{D}^T_{\pi}x^{*}_{\pi,i}\frac{\partial J_L\left(x^*_{\pi},\pi\right)}{\partial x^{*}_{\pi,i}}\,
\end{equation}
before updating its decision via~\eqref{eq:pgd2}. To obtain individual $\textbf{D}_{\pi}x^{*}_{\pi,i}$, we leverage the fact that the computed lower-level NE has to solve the best-response optimization problem of the corresponding follower. Namely, to tackle the requirements of the Implicit Function Theorem, for every $i\in\mc I$, we formulate an optimization problem equivalent to~\eqref{eq:best_resp} and directly apply the theorem on the problem's KKT mapping $l_i\left(z_i,\pi\mid x_{-i}^*\right)$. To ease the notation in the following sections, we will suppress the subscript denoting dependence on $\pi$ when it is clear from the context and refer to the Jacobian of follower $i\in\mc I$ as $\textbf{D}_{\pi}x_{i}^*$. 
\subsection{Differentiating the KKT conditions}
For a given $\pi\in\mc P$ and the corresponding unique v-NE $x^*\in\mc X(\pi)$ of the lower-level game, the Implicit Function Theorem allows us to locally compute Jacobians $\textbf{D}_{\pi}x_{i}^*$ by applying the theorem on the KKT mapping $l_i\left(z_i,\pi\mid x_{-i}^*\right)$. For every $i\in\mc I$, let the set-valued map $\Xi_{i}^*:\mc P\rightrightarrows \mc Z_i$, with $\mc Z_i\defineas\R^{m_F}\times\R_{\geq0}^{m_{\operatorname{inq},i}}\times\R^{m_{\operatorname{eq},i}}$, be 
\begin{equation}
\label{eq:solmap}
    \Xi^{*}_i(\pi)\defineas\left\{z_i\in\mc Z_i\left|l_i\left(z_i,\pi\mid x_{-i}^*\right)=0\right.\right\}\,.
\end{equation}
Moreover, let $\Theta=\bigl[1,m_{\operatorname{inq},i}\bigr]\cap\N$, and the set of non-strongly active inequality constraints $\Gamma^\pi_{i}(x_{i}^*,\lambda_{i}^*)$ be
\begin{equation}
    \label{eq:set_ass3}
    \Gamma^\pi_{i}(x_{i}^*,\lambda_{i}^*)\defineas\left\{j\in\Theta\mid \lambda^{j*}_i=0 \land \boldsymbol{g}_i^{\operatorname{inq}}(x_{i}^*,\pi)_j=0\right\}\,,
\end{equation}
where $\boldsymbol{g}_i^{\operatorname{inq}}(x_i,\pi)_j$ is the $j$-th inequality constraint and $\lambda^{j*}_i$ represents the corresponding dual variable of the best-response optimization problem. With a slight abuse of notation, the Implicit Function Theorem from~\cite{Implicit} adapted to our problem reads as the following theorem.
\begin{theorem}[Theorem 1.B1 of~\cite{Implicit}]\label{th:1}
    Let Standing Assumptions~\ref{sass:1}--~\ref{sass:4} hold and $x^*\in\mc X(\pi)$ be the unique NE of the game $G^0(\mc I;\pi)$ for some $\pi\in\mc P$. Furthermore, let the best-response optimization problem of each agent $i\in\mc I$ be defined via~\eqref{eq:best_resp}, its KKT mapping $l_i(z_i,\pi\mid x_{-i}^*)$ via~\eqref{eq:KKToperator}, and 
     $\Xi^{*}_i(\pi)$ be defined via~\eqref{eq:solmap}. If $l_i(\hat{z}_i,\pi\mid x_{-i}^*)=0$, $\Gamma^\pi_{i}(\hat{x}_{i},\hat{\lambda}_{i})$ is empty and $\operatorname{\mathbf{D}}_{z_i}l_i(\hat{z}_i,\pi\mid x_{-i}^*)$ is non-singular for some $\hat{z}_i$, then the solution mapping $\Xi_{i}^*(\pi)$ has a single-valued localization $z_{i}^*$ around $\hat{z}_i=(\hat{x}_i,\hat{\lambda}_i,\hat{\nu}_i)$, that is continuously differentiable in a neighbourhood $\Omega_{\pi}$ of $\pi$, with the Jacobian satisfying for every $\pi\in\Omega_{\pi}$ 
    \begin{equation}
        \label{eq:ImplicitJacobian}
        \operatorname{\mathbf{D}}_{\pi}z^{*}_i\left(\pi\right)=-\operatorname{\mathbf{D}}_{z_i}^{-1}l_i\left(\hat{z}_i,\pi\mid x_{-i}^*\right)\operatorname{\mathbf{D}}_{\pi}l_i\left(\hat{z}_i,\pi\mid x_{-i}^*\right)\,,
    \end{equation}
    where $\operatorname{\mathbf{D}}_{z_i}l_i\left(\hat{z}_i,\pi\mid x_{-i}^*\right)$ and $\operatorname{\mathbf{D}}_{\pi}l_i\left(\hat{z}_i,\pi\mid x_{-i}^*\right)$ satisfy
    \begin{equation} \label{eq:matinv}       \operatorname{\mathbf{D}}_{z_i}l_i\defineas\left[\begin{array}{ccc}\operatorname{\mathbf{D}}_{x_i}\nabla_{x_i}\mc L_i, & \operatorname{\mathbf{D}}_{x_i}^T\boldsymbol{g}_i^{\operatorname{inq}}, & \operatorname{\mathbf{D}}_{x_i}^T\boldsymbol{g}_i^{\operatorname{eq}} \\ \operatorname{Dg}\bigl(\hat{\lambda}_i\bigr)\operatorname{\mathbf{D}}_{x_i}\boldsymbol{g}_i^{\operatorname{inq}}, & \operatorname{Dg}\bigl(\boldsymbol{g}_i^{\operatorname{inq}}\bigr), & \operatorname{\mathbf{0}}\\
        \operatorname{\mathbf{D}}_{x_i}\boldsymbol{g}_i^{\operatorname{eq}}, & \operatorname{\mathbf{0}}, & \operatorname{\mathbf{0}}\end{array}\right]
    \end{equation}
    \begin{equation}     \operatorname{\mathbf{D}}_{\pi}l_i\defineas\left[\begin{array}{c}\operatorname{\mathbf{D}}_{\pi}\nabla_{x_i}\mc L_i\\ \operatorname{Dg}\bigl(\hat{\lambda}_i\bigr)\operatorname{\mathbf{D}}_{\pi}\boldsymbol{g}_i^{\operatorname{inq}}\\
        \operatorname{\mathbf{D}}_{\pi}\boldsymbol{g}_i^{\operatorname{inq}}\end{array}\right]\,.
    \end{equation}
    \normalsize
\end{theorem}
The triplet $\hat{z}_i=(x_{i}^*,\lambda_{i}^*,\nu_{i}^*)$, with $x_{i}^*$, $\lambda_{i}^*$ and $\nu_{i}^*$ being the solution of $G^0_i(\pi,x_{-i}^*)=\min_{x_i\in\mc X_i(\pi)}J_i(x_i,x_{-i}^*,\pi)$, satisfies $l_i(\hat{z}_i,\pi\mid x_{-i}^*)=0$. However, based on the Implicit Function Theorem, extracting the derivative $\operatorname{\mathbf{D}}_{\pi}x_{i}^*$ from $\operatorname{\mathbf{D}}_{\pi}z_{i}^*$ requires that the matrix $\operatorname{\mathbf{D}}_{z_i}l_i(\hat{z}_i,\pi\mid x_{-i}^*)$ be invertible. The necessary condition for this to hold is that the set $\Gamma_i^\pi(x_{i}^*,\lambda_{i}^*)$ be empty. Namely, observe that $\Gamma_i^\pi(x_{i}^*,\lambda_{i}^*)\neq\emptyset$ implies that there would exist a zero row in $\operatorname{\mathbf{D}}_{z_i}l_i(\hat{z}_i,\pi \mid x_{-i}^*)$, hence making it singular. On the other hand, the sufficient condition for the Implicit Function Theorem to hold directly depends on the structure of the game resulting from the nature of the application and the NE computed prior to the leader's strategy update step. To ensure this, in Section~\ref{subsec:3}, we reorganize the constraints of the original best-response optimization problem to form an equivalent one whose KKT map $l_i(z_i,\pi\mid x_{-i}^*)$ yields invertible $\operatorname{\mathbf{D}}_{z_i}l_i({z}_i,\pi \mid x_{-i}^*)$.

\subsection{Equivalent best-response optimization problem}\label{subsec:3}
For a particular $\pi\in\mc P$, let the unique v-NE of the lower-level game $G^0(\mc I;\pi)$ be $x^*\in\mc X(\pi)$. For every follower $i\in\mc I$, its NE strategy $x^{*}_i\in\mc X_i(\pi)$ explicitly provides information on what inequality constraints are active for a particular leader's strategy. Let $\mc A_i(x^{*}_i)$ represent the set of all active inequality constraints at $x^{*}_i$, i.e.,
\begin{equation}
\label{eq:ax}
    \mc A_i\left(x^{*}_i\right)\defineas\bigl\{j\in\bigl[1,m_{\operatorname{inq},i}\bigr]\cap\N\mid\boldsymbol{g}_i^{\operatorname{inq}}(x^{*}_i,\pi)_j=0\bigr\} \,.
\end{equation}
Consequently, let the complement of $\mc A_i\left(x_{i}^*\right)$ be 
\begin{equation}
\label{eq:axplus}
    \mc A^{\dagger}_i\left(x_{i}^*\right)=\left(\left[1,m_{\operatorname{inq},i}\right]\cap\N\right)\setminus\mc A_i\left(x_{i}^*\right)\,.
\end{equation}
If $\left|\mc A_i\left(x_{i}^*\right)\right|=m_{\text{act},i}> 0$, then we can define 
\begin{equation}\label{eq:gupper}
    \boldsymbol{\overline{g}}^{\operatorname{inq}}_i(x_i,\pi)=\text{col}\Big((\boldsymbol{g}^{\operatorname{inq}}_i(x_i,\pi)_j)_{j\in\mc A_i\left(x_{i}^*\right)}\Big)\,,
\end{equation}
\begin{equation}\label{eq:glower}
    \boldsymbol{\underline{g}}^{\operatorname{inq}}_i(x_i,\pi)=\text{col}\Big((\boldsymbol{g}^{\operatorname{inq}}_i(x_i,\pi)_j)_{j\in\mc A_i^{\dagger}\left(x_{i}^*\right)}\Big)\,,
\end{equation}
that effectively split the inequality constraints into a set of active and inactive ones. This allows us to formulate an auxiliary best-response optimization problem equivalent to~\eqref{eq:best_resp} whose corresponding set $\Gamma_i^\pi(x_{i}^*,\lambda_{i}^*)$ is empty.
\begin{lemma}\label{lem:2}
   Let Standing Assumptions~\ref{sass:1}--~\ref{sass:4} hold and $x^*\in\mc X(\pi)$ be the unique NE of the game $G^0(\mc I;\pi)$ for some $\pi\in\mc P$. Moreover, let $\mc A_i(x^{*}_i)$ and $A_i^{\dagger}(x^{*}_i)$ be defined as~\eqref{eq:ax} and~\eqref{eq:axplus} and $|\mc A_i(x^{*}_i)|=m_{\operatorname{act},i}\neq 0$. If $\boldsymbol{\overline{g}}^{\operatorname{inq}}_i(x_i,\pi)$ and $\boldsymbol{\underline{g}}^{\operatorname{inq}}_i(x_i,\pi)$ are given by~\eqref{eq:gupper} and~\eqref{eq:glower}, then $x^{*}_i\in\mc X_i(\pi)$ solves the best-response problem $G^0_i(\pi,x^{*}_{-i})$ given by~\eqref{eq:best_resp} if and only if it solves the surrogate problem 
   \begin{equation}
    \overline{G}^0_{i}\left(\pi,x^{*}_{-i}\right):=\left\{\begin{array}{c}
    \displaystyle\min_{x_i\in\R^{m_F}}J_i\left(x_i,x^{*}_{-i},\pi\right) \\
    \operatorname{s.t. }\boldsymbol{\underline{g}}^{\operatorname{inq}}_i(x_i,\pi)\leq \mathbf{0}_{m_{\operatorname{inq},i}-m_{\operatorname{act},i}}\\
    \quad\boldsymbol{\overline{g}}^{\operatorname{eq}}_i(x_i,\pi)=\mathbf{0}_{m_{\operatorname{eq},i}+m_{\operatorname{act},i}} 
    \end{array}\right\} \,,
    \label{eq:best_resp2}
\end{equation}
where $\boldsymbol{\overline{g}}_i^{\operatorname{eq}}(x_i,\pi)=\bigl[{\boldsymbol{g}_i^{\operatorname{eq}}}^T(x_i,\pi),\:\:  {\boldsymbol{\overline{g}}_i^{\operatorname{inq}}}^T(x_i,\pi)\bigr]^T$.
\end{lemma}
\begin{pf}
Observe that both problems are convex, so it suffices to look at their KKT optimality conditions. If $x_{i}^*\in\mc X_i(\pi)$ solves~\eqref{eq:best_resp} for some $\pi$, then $\nabla_{x_i}[J_i(x_{i},x_{-i}^*,\pi)+\lambda_{i}^T\boldsymbol{g}_i^{\operatorname{inq}}(x_{i},\pi)+\nu_{i}^T\boldsymbol{g}_i^{\operatorname{eq}}(x_{i},\pi)]=0$ is satisfied for $x_{i}^*$ and some feasible $\lambda_{i}^*$ and $\nu_{i}^*$. We can partition $\lambda^*$ into $\overline{\lambda}_i$ and $\underline{\lambda}_i$ and rewrite $\nabla_{x_i}[J_i(x_{i},x_{-i}^*,\pi)+\underline{\lambda}_{i}^T\boldsymbol{\underline{g}}_i^{\operatorname{inq}}(x_i,\pi)+\nu_{i}^T\boldsymbol{g}_i^{\operatorname{eq}}(x_{i},\pi)+\overline{\lambda}_{i}^T\boldsymbol{\overline{g}}_i^{\operatorname{inq}}(x_{i},\pi)]=0$. However, this is exactly the KKT stationarity condition of the surrogate best-response problem~\eqref{eq:best_resp2} for $\overline{\nu}_i^T=[\nu_i^T,\:\overline{\lambda}_i^T]$. Since the primal and dual feasibility conditions are equivalent, the proof is completed.\qed  
\end{pf}
We can now postulate the following results regarding the diferentiability of the KKT mapping $l_i\left(z_i,\pi\mid x_{-i}^*\right)$.
\begin{theorem}\label{th:2}
 Let Standing Assumptions~\ref{sass:1}--~\ref{sass:4} hold and $x^*\in\mc X(\pi)$ be the unique NE of the game $G^0(\mc I;\pi)$ for some $\pi\in\mc P$. Let the auxiliary best-response optimization problem $\overline{G}^0_{i}(\pi,x_{-i}^*)$ be defined as in Lemma~\ref{lem:2}, $\hat{z}_i=(x_{i}^*,\lambda_{i}^*,\nu_{i}^*)$ be its solution and $\operatorname{\mathbf{D}}_{z_i}l_i(\hat{z}_i,\pi\mid x_{-i}^*)$ be defined as in Theorem~\ref{th:1}. If $\operatorname{\mathbf{D}}_{x_i}\nabla_{x_i}\mc L_i(\hat{z}_i,\pi)\succ0$ and $\boldsymbol{\operatorname{D}}_{x_i}\boldsymbol{\overline{g}}_i^{\operatorname{eq}}(x_i,\pi)$ has full row rank for $\hat{x}_i$, then the matrix $\operatorname{\mathbf{D}}_{z_i}l_i(\hat{z}_i,\pi\mid x_{-i}^*)$ associated with $\overline{G}^0_{i}(\pi,x_{-i}^*)$ is invertible and the Jacobian $\operatorname{\mathbf{D}}_{\pi}x_{i}^*$ is given by 
    \begin{equation}
        \operatorname{\mathbf{D}}_{\pi}x_{i}^*=-\Sigma_1^{-1}\left[\Sigma_3-\Sigma_2^T(\Sigma_2\Sigma_1^{-1}\Sigma_2^T)^{-1}(\Sigma_3-\Sigma_4)\right]\,,
    \end{equation}
    where $\Sigma_1=\operatorname{\mathbf{D}}_{x_i}\nabla_{x_i}\mc L_i$, $\Sigma_2=\operatorname{\mathbf{D}}_{x_i}\boldsymbol{g}_i^{\operatorname{eq}}$, $\Sigma_3=\operatorname{\mathbf{D}}_{\pi}\nabla_{x_i}\mc L_i$ and $\Sigma_4=\operatorname{\mathbf{D}}_{\pi}\boldsymbol{g}_i^{\operatorname{eq}}$ are all evaluated at $\hat{z}_i,\pi,x_{-i}^*$.
\end{theorem}
\begin{pf}
We start by noting that $\boldsymbol{\underline{g}}_i^{\operatorname{inq}}(x_{i}^*,\pi)<0$ guarantees that $\lambda_{i}^*=\mathbf{0}$ due to complementary slackness, and hence $\overline{\Gamma}^{\pi}_{i}=\emptyset$. In order to prove invertibility of $\operatorname{\mathbf{D}}_{z_i}l_i(\hat{z}_i,\pi\mid x_{-i}^*)$, we invoke Lemma~\ref{lem:3} listed in Appendix. Namely, we can partition $\operatorname{\mathbf{D}}_{z_i}l_i(\hat{z}_i,\pi\mid x_{-i}^*)$ into blocks $M_1$, $M_2$, $M_3$ and $M_4$, evaluated at $\hat{z}_i,\pi$ and $x_{-i}^*$, such that
 $M_1=\operatorname{\mathbf{D}}_{x_i}\nabla_{x_i}\mc L_i(z_i,\pi)$,
\begin{equation}
    \label{eq:part23}
    M_2^T=\left[\begin{array}{c}\operatorname{\mathbf{D}}_{x_i}\boldsymbol{\underline{g}}_i^{\operatorname{inq}}(x_i,\pi)\\ \operatorname{\mathbf{D}}_{x_i}\boldsymbol{\overline{g}}_i^{\operatorname{eq}}(x_i,\pi)\end{array}\right],\quad M_3=\left[\begin{array}{c}\mathbf{0}\\ \operatorname{\mathbf{D}}_{x_i}\boldsymbol{\overline{g}}_i^{\operatorname{eq}}(x_i,\pi)\end{array}\right]\,,
\end{equation}
\begin{equation}
    \label{eq:part4}
    M_4=\left[\begin{array}{cc}\operatorname{Dg}(\boldsymbol{\underline{g}}_i^{\operatorname{inq}}(x_i,\pi)) & \mathbf{0} \\
    \mathbf{0} & \mathbf{0}\end{array}\right]\,.
\end{equation} 
For $\hat{z}_i,\pi,x_{-i}^*$, the Shur complement of $M_1$ is given by
\begin{equation}
    \operatorname{Sh}\left(M_1\right)\defineas\left[\begin{array}{cc}\operatorname{Dg}(\boldsymbol{\underline{g}}_i^{\operatorname{inq}}) & \mathbf{0} \\ \star & -\operatorname{\mathbf{D}}_{x_i}\boldsymbol{\overline{g}}_i^{\operatorname{eq}}M_1^{-1}\operatorname{\mathbf{D}}_{x_i}^T\boldsymbol{\overline{g}}_i^{\operatorname{eq}}\end{array}\right]\,.
\end{equation}
Since $\boldsymbol{\underline{g}}_i^{\operatorname{inq}}(x_{i}^*,\pi)$ encompasses inactive inequality constraints, we have that $\operatorname{Dg}(\boldsymbol{\underline{g}}_i^{\operatorname{inq}}(x_{i}^*,\pi))\prec0$. Similarly, because $\operatorname{\mathbf{D}}_{x_i}\nabla_{x_i}\mc L_i(\hat{z}_i,\pi)\succ0$ and $\boldsymbol{\operatorname{D}}_{x_i}\boldsymbol{\overline{g}}_i^{\operatorname{eq}}(x_i,\pi)$ has full row rank, we have that $\operatorname{\mathbf{D}}_{x_i}\boldsymbol{\overline{g}}_i^{\operatorname{eq}}M_1^{-1}\operatorname{\mathbf{D}}_{x_i}^T\boldsymbol{\overline{g}}_i^{\operatorname{eq}}\succ 0$, making $\operatorname{Sh}(M_1)$, and hence $\operatorname{\mathbf{D}}_{z_i}l_i(\hat{z}_i,\pi\mid x_{-i}^*)$, nonsingular. Moreover, based on Theorem~\ref{th:1}, we have $\operatorname{\mathbf{D}}_{\pi}x_{i}^*=-\overline{M}_1\Sigma_3-\overline{M}_2[\mathbf{0}^T,\Sigma_4^T]^T$, where Lemma~\ref{lem:3} gives $\overline{M}_1=\Sigma_1^{-1}(\mathbb{I}-\Sigma_2^T(\Sigma_2\Sigma_1^{-1}\Sigma_2^T)^{-1}\Sigma_2\Sigma_1^{-1})$ and $\overline{M}_2=-\Sigma_1^{-1}\left[\star,\:\:-\Sigma_3^T(\Sigma_3\Sigma_1^{-1}\Sigma_3^T)^{-1}\right]$. Direct computation of the right-hand side completes the proof. \qed
\end{pf}
Theorem~\ref{th:2} offers two general conditions that can be used to assess the invertibility of $\operatorname{\mathbf{D}}_{z_i}l_i(\hat{z}_i,\pi\mid x_{-i}^*)$ and, consequently, establish the well-posedness of the Jacobian $\boldsymbol{\operatorname{D}}_{\pi}x_{i}^*$. This essentially involves confirming the typical structural characteristics of the followers' cost functions and constraint sets for commonly encountered instances of Stackelberg games. On the other hand, the closed form of the Jacobian is a direct consequence of Lemma~\ref{lem:3} and shows that the Jacobian retains constant functional form during the segments of the leader's update procedure with the same sets of active inequality constraints. In the following section, we will further discuss the applicability of Theorem~\ref{th:2} in particular cases. However, before we proceed, we will first present the formal convergence guarantees for the more general case.
\begin{theorem}\label{th:3}
    Let the Stackelberg game be defined as~\eqref{eq:SG} under Standing Assumptions~\ref{sass:1}--~\ref{sass:4}. At every update step $t\in\N$ of the leader, let $x^*_t\in\mc X(\pi_t)$ be the unique v-NE of the lower level game and the surrogate best-response optimization problem of the $i$-th follower be defined as in Lemma~\ref{lem:2}. If the sequence $\{\pi_t\}$ generated by the projected gradient descent method defined by~\eqref{eq:pgd3} and~\eqref{eq:pgd2} fulfills the conditions of Theorem~\ref{th:2}, then it holds that 
    \begin{equation}
    \label{eq:lim}
        \lim_{t\rightarrow+\infty}\left[J_L\bigl(x^{*}_{\pi_{t+1}},\pi_{t+1}\bigr)-J_L\bigl(x^{*}_{\pi_{t}},\pi_t\bigr)\right]=0\,,
    \end{equation}
    and every limit point of $\left\{\pi_t\right\}$ is stationary.
\end{theorem}
\begin{pf}
First, note that based on the Armijo rule, the sequence $\{J_L(x^*_{\pi_t},\pi_t)\}_{t=1}^{\infty}$ is monotonically nonincreasing. Because $J_L\left(x^*_{\pi},\pi\right)$ is continuous in $z^T=[(x^*_\pi)^T,\pi^T]$ and $\cup_{\pi\in\mc P}\mc X(\pi)\times\mc P$ is compact, there exists $J_L^{\text{min}}\in\R$ such that $J_L\left(x^*_\pi,\pi\right)\geq J_L^{\text{min}}$ for all $z\in\cup_{\pi\in\mc P}\mc X(\pi)\times\mc P$. Since the sequence $\left\{J_L\left(x^*_{\pi_t},\pi_t\right)\right\}_{t=1}^{\infty}$ is monotonically nonincreasing and bounded, it converges to a finite value implying $\lim_{t\rightarrow +\infty}[J_L(x^{*}_{\pi_{t+1}},\pi_{t+1})-J_L\left(x^*_{\pi_t},\pi_t\right)]=0$. Since Theorem~\ref{th:2} guarantees that $J_L\left(x^{*}_{\pi},\pi\right)$ is continuously differentiable at every $\pi\in\mc P$, every limit point of $\left\{\pi_t\right\}$ is stationary based on~\cite[P2.3.3]{Bertsekas/99}. \qed
\end{pf}
\begin{algorithm}[tbp]
    \caption{Finding leader's optimal strategy}\label{al:1}
    \begin{algorithmic}[1]
    \State \textbf{Input:}  $\gamma$, $\beta$, $\overline{s}$, $\delta$, $\varepsilon$, $T$
    \State \textbf{Output:} $\pi$
    \State $\pi_0=\text{Initialize}()$;
    \For {$t\gets 0 \textbf{ to } T$}
        \State $x^*_{\pi_t}=\text{ComputeVariationalNE}(\pi_t)$;
        \For {$i\in\mc I$}\Comment{In parallel}
            \State $\text{Define\:} \boldsymbol{\underline{g}}_i^{\operatorname{inq}}(x_{\pi_t,i},\pi_t),\: \boldsymbol{\overline{g}}_i^{\operatorname{eq}}(x_{\pi_t,i},\pi_t)$;
            \State Obtain $\boldsymbol{\operatorname{D}}_{\pi}x_{\pi_t,i}^*$ using Theorem 1 on $\overline{G}^0_{i}$;
        \EndFor
        \State \textbf{Leader}:
        \State \quad\:\:$\frac{\text{d} J_L\left(\cdot\right)}{\text{d} \pi}=\frac{\partial J_L\left(\cdot\right)}{\partial \pi}+\sum_{i\in\mc I}\textbf{D}^T_{\pi}x^{*}_i\frac{\partial J^L\left(\cdot\right)}{\partial x^{*}_i}$;
        \State \quad\:\:$s_t=\text{ArmijoStep}\left(\beta, \overline{s},\delta,\pi_t,\frac{\text{d} J^L\left(\cdot\right)}{\text{d} \pi}\right)$;
        \State \quad\:\:$\pi_{t+1}=\pi^+\left(\pi_t,s_t\right)$;
     \EndFor
    \end{algorithmic}
\end{algorithm}
The complete iterative procedure for finding a local Stackelberg equilibrium is outlined in Algorithm~\ref{al:1}. As previously shown, the Jacobian computations required for performing the update step on the upper level can be entirely parallelized regardless of the type of game being played among the followers. Therefore, the overall distributed nature of the complete procedure is entirely dictated by the subproblem of computing the v-NE for a particular leader's strategy. In the forthcoming Section 4, we narrow our focus to a special case of games presented in Defintion~\ref{ex:1} with polytopic constraints of the followers. As demonstrated in~\cite{Hierarchical, ecc2023, Paccagnan2016a, Paccagnan2019}, these games enable the computation of the followers' v-NE in a semi-decentralized manner, involving only the exchange of aggregated follower decisions facilitated by a central aggregator entity, i.e., the communication hub. When $\mc P$ is a polytope, we demonstrate how this communication hub can also be used to design a decentralized warm-start procedure that may help mitigate issues arising from an unfavorable initial value of the leader's decision variable $\pi_0\in\mc P$. 
\section{Quadratic aggregative Stackelberg games}\label{sec:LQG}
We focus on a particular instance of games in Definition~\ref{ex:1} with $\pi_i=\pi_j=\pi$ for all $i,j\in\mc I$. In that case, the lower-level agents minimize a quadratic cost of the form
\begin{equation}\label{eq:costq}
    J_i=\frac{1}{2}x_i^TP_ix_i+x_i^TQ_i\sigma(x_{-i})+r_i^Tx_i+x_i^TS_i\pi\,,
\end{equation}
under $\pi$-parametrized local polytopic constraints $x_i\in\mc X_i(\pi)$ given by $\boldsymbol{g}_i^{\operatorname{inq}}(x_i,\pi)=G_i(\pi)x_i-b_i(\pi)$ and $\boldsymbol{g}_i^{\operatorname{eq}}(x_i,\pi)=A_i(\pi)x_i-b_i(\pi)$. In light of Theorem~\ref{th:2}, if $x^*\in\mc X(\pi)$ denotes the v-NE of the lower level game and $|\mc A_i(x_{i}^*)|\neq 0$, we proceed to formulate the surrogate best-response optimization problem by letting $\overline{G}_i(\pi)\in\R^{m_{\operatorname{act},i}\times m_F}$ be a matrix whose rows are the rows of $G_i(\pi)$ listed in $\mc A_i(x_{i}^*)$. Moreover, we let $\underline{G}_i(\pi)$ encompass all the remaining rows of $G_i(\pi)$ and decompose the vector $h_i(\pi)$ into $\overline{h}_i(\pi)$ and $\underline {h}_i(\pi)$ such that $\overline{G}_i(\pi)x_{i}^*=\overline{h}_i(\pi)$ and $\underline{G}_i(\pi)x_{i}^*<\underline{h}_i(\pi)$. For the cost~\eqref{eq:costq}, the conditions of Theorem~\ref{th:2} ensuring the invertibility of 
    \begin{equation}
    \label{eq:Kbar}
\textbf{D}_{z_i}l_i=\left[\begin{array}{c:cc}
    P_i & \underline{G}_i^T(\pi) & \overline{A}_i^T(\pi) \\
    \hdashline
    \mathbf{0} & \operatorname{Dg}\left(\underline{G}_i(\pi)x_{i}^*-\underline{h}_i(\pi)\right) & \mathbf{0} \\
    \overline{A}_i(\pi) & \mathbf{0} & \mathbf{0}
    \end{array}\right]\,,    
    \end{equation}
related to $\overline{G}^0_i(\pi,x_{-i}^*)$ reduce to $\operatorname{\mathbf{D}}_{x_i}\nabla_{x_i}L(\hat{z}_i,\pi)=P_i\succ0$ and ensuring that $\overline{A}_i(\pi)=[A_i^T(\pi),\: \overline{G}_i^T(\pi)]^T$ is full row rank. Note that the latter can be easily accounted for during the construction step of the surrogate best-response problem. When designing $\overline{A}_i(\pi)$, it suffices to exclude the active inequality constraints for which there already exists a linearly dependant equality constraint.

\begin{figure}
    \centering
        \begin{tikzpicture}[scale=1.1]
        \pgfmathtruncatemacro{\numEllipses}{4}
        
        \def\majorRadius{3}
        \def\minorRadius{2}
        \def\minorRadiusIncrement{0.4}
        \def\majorRadiusIncrement{0.6}
        
        \foreach \i in {1,...,\numEllipses} {
            \draw[rotate=25, fill=blue!20, fill opacity=0.8/\numEllipses*\i, draw opacity=0.1] (0,0) ellipse ({\majorRadius - \i*\majorRadiusIncrement} and {\minorRadius - \i*\minorRadiusIncrement});
        }

        \draw[dashed](-1.8,-1.6) -- (-0.8,1.6);
        \draw[dashed](-1.8,-1.6) -- (0.8,-1.6);
        \draw[dashed](0.8, -1.6) -- (2.4, 0.0);
        
        \node[scale=0.6](n1) at (-0.5,-1.75) {$(g_{i,j+1}^1)^Tx_i=h_{i,j+1}$};
        \node[rotate=75, scale=0.6](n2) at (-1.7, -0.8) {$(g_{i,j}^1)^Tx_i=h_{i,j}$};
        \node[rotate=45, scale=0.6](n7) at (1.7, -0.9) {$(g_{i,j+2}^1)^Tx_i=h_{i,j+2}$};
        
        \node[draw=red, circle, fill=red, inner sep=0.8pt](n3) at (-1.3, 0) {};

        \node[draw=red, circle, fill=red, inner sep=0.8pt](n4) at (-0.8,1.6) {}; 

        \draw[->, red](n4) -- (n3);
        \node[red, scale=0.6](n5) at (-0.8,1.8) {$x_i^*\left(\pi_0^a\right)$};
        \node[red, scale=0.6](n6) at (-1.7,0.2) {$x_i^*\left(\pi_k^a\right)$};

        \node[](n8) at (0.6,-1.3) {$\mc X_i$};

        \node[draw=orange, circle, fill=orange, inner sep=0.8pt](n9) at (1.5, 1.5) {};
        \node[draw=orange, circle, fill=orange, inner sep=0.8pt](n10) at (0,0) {};

        \draw[->, orange] (n9) to[out=-45, in=45] (n10);

        \node[orange, scale=0.6](n11) at (1.8, 1.8) {$x_i^*\left(\pi_0^b\right)$};
        \node[orange, scale=0.6](n12) at (0.0, -0.2) {$x_i^*\left(\pi_k^b\right)$};

        \node[](n13) at (0,0) {};
        
    \end{tikzpicture}
    \caption{Ilustrative example of the lower-level NE evolution when the projected gradient descent algorithm is initialized by $\pi_0^a$, i.e., such that the NE is on the boundary of the feasible set $\mc X_i$ (red), and by $\pi_0^b$, i.e., such that the NE is in the interior (orange).}
    \label{fig:exem}
\end{figure}
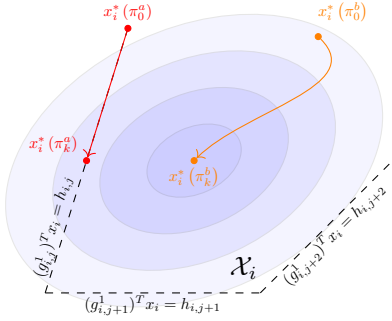

As mentioned earlier, quadratic aggregative games allow us to tackle the initialization problem to a certain extent. Our previous work~\cite{ecc2023} empirically demonstrated that by adjusting the $\pi_0\in\mc P$ value used to initiate the iterative update procedure of the leader we can give rise to local Stackelberg equilibria of varying quality with respect to the value of the leader's objective. While it is possible to sample the leader's action space and repeat the complete iterative process multiple times to find a better solution, opting for initial values $\pi_0$ that immediately lead to a v-NE that causes certain inequality constraints to become active can unnecessarily hinder the subsequent steps of the procedure. As illustrated in Figure~\ref{fig:exem}, since the Jacobian is calculated with respect to the surrogate best-response problem, poorly choosing the initial value $\pi_0$ means that we effectively start with a lower-level game where each follower has more equality constraints than originally postulated. Although later on we do not have precise control over the trajectory of the leader's iterative procedure, if $\mc P$ and $\mc X_i(\pi)$ are polytopes, with $\mc X_i(\pi)$ being implicitly governed by a linear map of $\pi$ and $x_i$, the structure of the analyzed games allows us to increase the flexibility of the algorithm by at least providing a feasible point $\pi_0$ that would yield an interior point v-NE in the first iteration of the algorithm. In other words, in the following section we will present a decentralized method for computing a $\pi_0$, should such a point exist, that results in a v-NE of the lower-level game that renders no inequality constraint active.

\subsection{Decentralized initialization procedure for agents with polytopic actions spaces}

As mentioned earlier, a decentralized initialization procedure can be devised for specific configurations of the agents' constraint sets. Before we go deeper into details about the procedure, we summarize all structural requirements on $\mc P$ and $\mc X_i(\pi)$ in the following assumption. 
\begin{assumption}\label{ass:s1}
    The constraint sets $\mc P\subset\R^{m_L}$ and $\mc X_i(\pi)\subset\R^{m_F}$ are given by bounded polytopes
\begin{equation}\label{eq:mcp}
    \mc P\defineas\{\pi \mid \operatorname{\mathbf{A}}_{\pi}\pi=\operatorname{\mathbf{b}}_{\pi}\land\operatorname{\mathbf{G}}_{\pi}\pi\leq\operatorname{\mathbf{h}}_{\pi}\}\,,
\end{equation}
\begin{equation}\label{eq:xipi}
    \mc X_i(\pi)\defineas\{x_i \mid A_ix_i+A_i^{\pi}\pi=b_i\land G_ix_i+G_i^{\pi}\pi\leq h_i\}\,.
\end{equation}
\end{assumption}
To find a $\pi_0\in\mc P$ that yields an interior point v-NE $x^*\in\mc X(\pi_0)$, 
we are essentially interested in ensuring that such a $\pi_0$ entails existence of a positive slack vector $\delta_i$ such that $G_ix^{*}_i+G_i^{\pi}\pi_0+\delta_i\leq h_i$. However, as we hope to avoid imposing information exchange regarding the personal constraint sets of the followers, we anticipate a consensus-based mechanism to compute $\pi_0$.

For every $i\in\mc I$, we look at the KKT conditions of the best-response optimization problem and define $\psi_i\in\R^{m_{\psi_i}}$, with $m_{\psi_i}=Nm_F+m_L+m_{\operatorname{eq},i}+m_{\operatorname{inq},i}$ and 
$$\psi_i\defineas[\tilde{\textbf{x}}_i^T,\:\tilde{\textbf{p}}_i^T,\:\nu_i^T,\:\delta_i^T]^T,$$
where $\tilde{\textbf{x}}_i=\text{col}((\tilde{x}_i^j)_{j\in\mc I})\in\mc X(\pi_0)$ denotes the $i$-th follower's local copy of the complete v-NE $x^*$, $\tilde{\textbf{p}}_i\in\mc P$ being the local copy of the vector $\pi_0$, $\nu_i$ being the Lagrangian multiplier associated with equality constraints of the best response optimization problem and $\delta_i$ is the slack vector that we are looking for. Furthermore, let $\Lambda_{\tilde{x}_i}$, $\Lambda_{\tilde{p}_i}$, $\Lambda_{\delta_i}$ and $\Lambda_i$ be selection matrices such that $\Lambda_{\tilde{x}_i}\psi_i=\tilde{x}^i_i$, $\Lambda_{\tilde{p}_i}\psi_i=\textbf{p}_i$, $\Lambda_{\delta_i}\psi_i=\delta_i$ and $\Lambda_i\psi_i=[\tilde{\textbf{x}}_i^T,\tilde{\textbf{p}}_i^T]^T$. We can now postulate a necessary and sufficient feasibility test based on linear programming. 
\begin{theorem}[Internal v-NE feasibility check]\label{th:4}
    Let the Stackelberg game be defined as~\eqref{eq:SG} under Standing Assumptions~\ref{sass:1}--~\ref{sass:4}, the structural Assumption~\ref{ass:s1} and objective functions given by~\eqref{eq:costq}. There exists a $\pi_0\in\mc P$ such that the corresponding v-NE, $x^*\in\mc X(\pi_0)$, of the lower-level game $G^0(\mc I;\pi_0)$ renders no inequality constraint active if and only if there exists  $\varepsilon>0$ such that the following linear optimization problem has a solution 
    \begin{mini!}
    {\{\psi_i\}_{i\in\mc I},\beta}{-\sum_{i\in\mc I}\mathbf{1}^T\Lambda_{\delta_i}\psi_i}
    {\label{mini:op1}}{}
    \addConstraint{\Lambda_i\psi_i-\beta=\mathbf{0}}{}{\label{eq:cc1}}
    \addConstraint{[W_i\:\:S_i^T\:\:A_i^T\:\:\mathbf{0}]\psi_i+r_i=\mathbf{0}}{}{\label{eq:cc2}}
    \addConstraint{(A_i\Lambda_{\tilde{x}_i}+A_i^{\pi}\Lambda_{\tilde{p}_i})\psi_i=b_i}{}{\label{eq:cc3}}
    \addConstraint{(G_i\Lambda_{\tilde{x}_i}+G_i^{\pi}\Lambda_{\tilde{p}_i}+\Lambda_{\delta_i})\psi_i\leq h_i}{}{\label{eq:cc4}}
    \addConstraint{-\Lambda_{\delta_i}\psi_i\leq-\varepsilon\mathbf{1}}{}{\label{eq:cc5}}
    \addConstraint{\operatorname{\mathbf{A}}_{\pi}\Lambda_{\tilde{\textbf{p}}_i}\psi_i=\operatorname{\mathbf{b}_{\pi}}}{}{\label{eq:cc6}}
    \addConstraint{\operatorname{\mathbf{G}}_{\pi}\Lambda_{\tilde{\textbf{p}}_i}\psi_i\leq\operatorname{\mathbf{h}_{\pi}}}{}{\label{eq:cc7}}
    \end{mini!}
where $W_i=[P_i,\quad\mathbf{1}^T_{N-1}\otimes \mathbb{I}]$.
\end{theorem}
\begin{pf}
    Firstly, observe that the dummy variable $\beta\in\R^{m_{\beta}}$, where $m_{\beta}=Nm_F+m_L$, ensures through~\eqref{eq:cc1} that all the followers have equal local copies of the pricing vector and the v-NE. Moreover, since $\varepsilon>0$,~\eqref{eq:cc4} and~\eqref{eq:cc5} ensure that any optimal solution of this problem renders no inequality constraint active. Due to complementarity slackness, we search for solutions where the dual variables satisfy $\lambda_i=\mathbf{0}$. Hence,~\eqref{eq:cc2} represents the stationarity condition of the KKT system for the convex best-response optimization problem. By adding~\eqref{eq:cc3} and~\eqref{eq:cc4}, we form a complete set of KKT optimality conditions, so any optimal $\beta$ corresponds to a $\pi_0$ for which the v-NE is an interior point. \qed
\end{pf}
Firstly, we note that~\eqref{mini:op1} could have been posed as a feasibility problem for simplicity. However, we opt for the proposed functional form as it places the attained v-NE further away from the boundaries. Clearly, the optimal solution $\beta^*$ encodes a viable $\pi_0$ and its corresponding v-NE, i.e., $\beta^*=[(x^*_{\pi_0})^T,\:\pi_0^T]$. Thanks to the separable objective function and constraints of the optimization problem~\eqref{mini:op1}, we can preserve the privacy of the followers and solve~\eqref{mini:op1} in a decentralized fashion through the consensus alternating direction method of multipliers (ADMM)~\cite{8186925}. Here, $\beta$ acts as a global variable to be shared among all the followers and is the only one that needs to be updated in a centralized manner, e.g., in case studies presented in~\cite{Hierarchical, ecc2023, Paccagnan2016a, Paccagnan2019}, this could be served by the same central entity required for computing the v-NE. If we let the polytope $$\Omega_{\psi_i}\defineas\{\psi_i\in\R^{m_{\psi_i}} \mid M_i^1\psi_i=v_i^1\land M_i^2\psi_i\leq v_i^2\}$$ encode all the constraints of~\eqref{mini:op1} except~\eqref{eq:cc1}, then the augmented Lagrangian of~\eqref{mini:op1} is
\begin{equation}\label{eq:lagp}
    \begin{split}
    \mc L_\rho(\{\psi_i\},\beta,\{y_i\})&=\sum_{i\in\mc I}-\mathbf{1}^T\Lambda_{\delta_i}\psi_i+I_{\Omega_{\psi_i}}(\psi_i)\\&+y_i^T(\Lambda_i\psi_i-\beta)+\frac{\rho}{2}\norm{\Lambda_i\psi_i-\beta}^2_2\,,
    \end{split}
\end{equation}
where $I_{\Omega_{\psi_i}}(\psi_i)$ denotes the indicator function and $\rho>0$ is an a priori chosen parameter. The consensus ADMM consists of repeating the following three steps
\begin{equation}\label{eq:consensus}
    \begin{array}{c}
         \psi_i^{k+1}=\displaystyle\argmin_{\psi_i\in\Omega_{\psi_i}} \mc L_\rho(\psi_i,\{\psi_j^k\}_{j\in\mc I\setminus\{i\}},\beta^k,\{y_i^k\}),  \\
         \beta^{k+1}=\displaystyle\argmin_{\beta\in\R^{m_\beta}}\mc L_\rho(\{\psi_i^{k+1}\},\beta,\{y_i^k\}), \\
         y_i^{k+1}=y_i^{k}+\rho(\Lambda_i\psi^{k+1}_i-\beta^{k+1}).
    \end{array}
\end{equation}
Due to the separability of the augmented Lagrangian, solving the $N$ convex quadratic optimization problems for updating individual $\psi_i$ can be done in parallel. The same holds for updating the dual variables $y_i$. On the other hand, the unconstrained quadratic minimization problem to be solved to obtain $\beta^{k+1}$ yields
\begin{equation}
    \beta^{k+1}=\frac{1}{N}\left[\frac{1}{\rho}\sum_{i\in\mc I}y_i^k+\sum_{i\in\mc I}\Lambda_i\psi_i^{k+1}\right]\,,
\end{equation}
and requires that the followers communicate their updated local estimate of $\pi_0$ and the corresponding v-NE, both encoded in $\psi^{k+1}_i$, to the central aggregator who will then update the consensus variable $\beta$. Formal convergence guarantees are given in the following theorem.
\begin{theorem}
    Let $\Delta^k_i=\Lambda_i\psi_i^k-\beta^k$ denote the residual at each iteration of the consensus ADMM given by~\eqref{eq:consensus}. If $A_i$ is full row rank then $\Delta_i^k\rightarrow 0$ when $k\rightarrow\infty$.
\end{theorem}
\begin{pf}
    We aim to directly invoke~\cite[Th 4.1]{admm}. Namely, for~\cite[Th 4.1]{admm} to hold, we first observe that the extended, real-valued function $\overline{f}=\sum_{i\in\mc I}-\mathbf{1}^T\Lambda_{\delta_i}\psi_i+I_{\Omega_{\psi_i}}(\psi_i)$ is closed, proper and convex. Secondly, we need to make sure that the solution set of~\eqref{mini:op1} is bounded. For this, it suffices to prove that $\Omega_{\psi_i}$ is bounded as then $\beta=\Lambda_i\psi_i$ is bounded as well. Under Standing Assumption~\ref{sass:1} $\tilde{\textbf{x}}_i\in\mc X(\pi)$ is bounded and $\tilde{\textbf{p}}_i\in\mc P$ is bounded because of~\eqref{eq:mcp}. From~\eqref{eq:cc2}, we have $A_i^T\nu_i=\gamma_i$ for $\gamma_i\defineas-r_i-W_i\tilde{\textbf{x}}_i-S_i^T\tilde{\textbf{p}}_i$. If $\gamma^j_i$ denotes the $j$-th element of the vector, then for every $j\in[1,m_F]\cap\N$ there exist $\underline{\gamma}_i^j,\overline{\gamma}_i^j\in\R$ such that $\underline{\gamma}_i^j\leq\gamma_i^j\leq\overline{\gamma}_i^j$ since both $\tilde{\textbf{x}}_i$ and $\tilde{\textbf{p}}_i$ are bounded. If we set $\gamma_i^{\operatorname{min}}=\min_j\underline{\gamma}_i^j$ and  $\gamma_i^{\operatorname{max}}=\max_j\overline{\gamma}_i^j$, then $\gamma_i^{\operatorname{min}}\mathbf{1}\leq A_i^T\nu_i\leq\gamma_i^{\operatorname{max}}\mathbf{1}$. If $A_i$ is full row rank, then the polytope $\gamma_i^{\operatorname{min}}\mathbf{1}\leq A_i^T\nu_i\leq\gamma_i^{\operatorname{max}}\mathbf{1}$ is bounded. Similarly, we can establish that $\delta_i$ is bounded based on~\eqref{eq:cc4} and~\eqref{eq:cc5}, which completes the proof.   \qed      
\end{pf}
In the following section, we will introduce in detail the two numerical case studies showcasing the performance of the main decentralized algorithm and its corresponding warm-start procedure. 

\definecolor{color0}{rgb}{0.83921568627451,0.152941176470588,0.156862745098039}
\definecolor{color1}{rgb}{1,0.498039215686275,0.0549019607843137}
\definecolor{color2}{rgb}{0.0901960784313725,0.745098039215686,0.811764705882353}
\begin {figure}
\centering
\resizebox{.45\textwidth}{!}{%
\begin{tikzpicture}[scale=0.75]

    \node (p1) at (0.5,0) {\includegraphics[width=.04\textwidth]{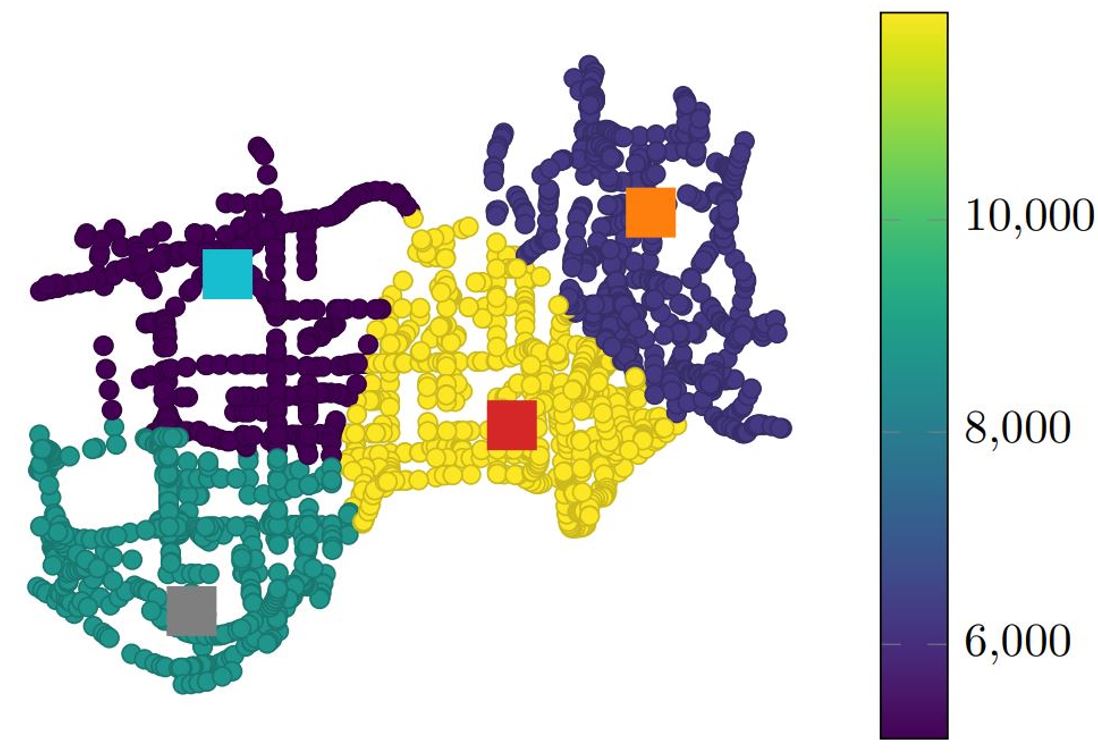}};
    \node (p2) at (0.5,-0.6) {\includegraphics[width=.04\textwidth]{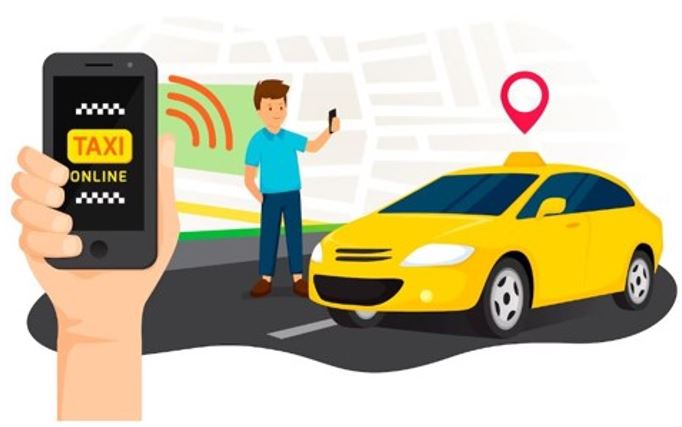}};
    \node[scale=0.09](n1) at (0.65, -0.18) {$x_i=N_i\left[\begin{array}{c}
         x_i^1  \\
         \vdots  \\
         x_i^4
    \end{array}\right]$};
    \node[rotate=90, scale=0.1, color=white](n2) at (0.817, 0.0) {Ride-hailing requests};
    
    \draw[rounded corners=1pt, line width=0.05pt] (0.02,0.35) rectangle (0.98,-0.9);

    \node (p3) at (-0.505,0.19) {\includegraphics[width=.015\textwidth]{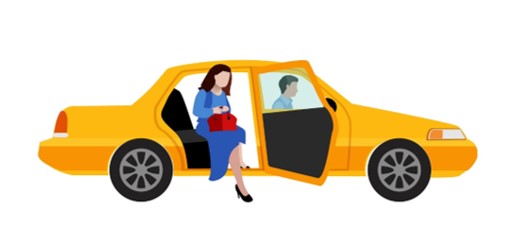}};
    \node (p4) at (-0.77,0.23) {\includegraphics[width=.01\textwidth]{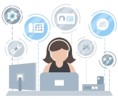}};
    \node[ scale=0.1](n3) at (-0.5, 0.28) {$N_1$ vehicles};
    \draw[rounded corners=1pt, line width=0.05pt] (-0.9,0.33) rectangle (-0.35, 0.12);
    \node [scale=0.15, cloud, draw, cloud puffs=10, cloud puff arc=120, aspect=2, inner ysep=0.01pt, minimum width=0.01pt, minimum height=0.01pt, line width=0.001pt, fill=white] (c1) at (-0.23,0.225) {$B_1$};
    \draw[line width=0.05pt](-0.35,0.225) -- (c1.west);
    \node[ scale=0.1](n33) at (-0.625, 0.36) {\textbf{Company} $ I_1$};
    
    \node (p5) at (-0.505,-0.31) {\includegraphics[width=.015\textwidth]{figures/taxi.jpg}};
    \node (p6) at (-0.77,-0.27) {\includegraphics[width=.01\textwidth]{figures/Operator.jpg}};
    \node[ scale=0.1](n4) at (-0.5, -0.22) {$N_2$ vehicles};
    \draw[rounded corners=1pt, line width=0.05pt] (-0.9,-0.17) rectangle (-0.35, -0.38);
    \node [scale=0.15, cloud, draw, cloud puffs=10, cloud puff arc=120, aspect=2, inner ysep=0.01pt, minimum width=0.01pt, minimum height=0.01pt, line width=0.001pt, fill=white] (c2) at (-0.23,-0.275) {$B_2$};
    \draw[line width=0.05pt](-0.35,-0.275) -- (c2.west);
    \node[ scale=0.1](n33) at (-0.625, -0.14) {\textbf{Company} $ I_2$};

    \node (p7) at (-0.505,-0.81) {\includegraphics[width=.015\textwidth]{figures/taxi.jpg}};
    \node (p8) at (-0.77,-0.77) {\includegraphics[width=.01\textwidth]{figures/Operator.jpg}};
    \node[ scale=0.1](n5) at (-0.5, -0.72) {$N_3$ vehicles};
    \draw[rounded corners=1pt, line width=0.05pt] (-0.9,-0.67) rectangle (-0.35, -0.88);
    \node [scale=0.15, cloud, draw, cloud puffs=10, cloud puff arc=120, aspect=2, inner ysep=0.01pt, minimum width=0.01pt, minimum height=0.01pt, line width=0.001pt, fill=white] (c3) at (-0.23,-0.775) {$B_3$};
    \draw[line width=0.05pt](-0.35,-0.775) -- (c3.west);
    \node[ scale=0.1](n33) at (-0.625, -0.64) {\textbf{Company} $ I_3$};

    \draw[-{Triangle[length=0.8pt, width=0.5pt]}, line width=0.05pt] (c1.east) -- (0.02, -0.275);
    \draw[-{Triangle[length=0.8pt, width=0.5pt]}, line width=0.05pt] (c2.east) -- (0.02, -0.275);
    \draw[-{Triangle[length=0.8pt, width=0.5pt]}, line width=0.05pt] (c3.east) -- (0.02, -0.275);

    \draw[rounded corners=1pt, line width=0.05pt] (-0.95, 0.4) rectangle (1.03, -0.95);

    \node (p9) at (-0.46, 0.725) {\includegraphics[width=.002\textwidth]{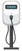}};
    \draw[fill, color=color0, line width=0.05pt] (-0.41, 0.75) rectangle (-0.34, 0.68);
    \node[color=color0,scale=0.1](n6) at (-0.41, 0.63) {Station $ M_1$};

    \node (p10) at (-0.16, 0.725) {\includegraphics[width=.002\textwidth]{figures/charger.jpg}};
    \draw[fill, color=color1, line width=0.05pt] (-0.11, 0.75) rectangle (-0.04, 0.68);
    \node[color=color1,scale=0.1](n6) at (-0.11, 0.63) {Station $ M_2$};

    \node (p11) at (0.14, 0.725) {\includegraphics[width=.002\textwidth]{figures/charger.jpg}};
    \draw[fill, color=white!49.8039215686275!black, line width=0.05pt] (0.19, 0.75) rectangle (0.26, 0.68);
    \node[color=white!49.8039215686275!black,scale=0.1](n6) at (0.19, 0.63) {Station $ M_3$};    

    \node (p12) at (0.44, 0.725) {\includegraphics[width=.002\textwidth]{figures/charger.jpg}};
    \draw[fill, color=color2, line width=0.05pt] (0.49, 0.75) rectangle (0.56, 0.68);
    \node[color=color2,scale=0.1](n6) at (0.49, 0.63) {Station $ M_4$};  

    \draw[rounded corners=1pt, line width=0.05pt] (-0.56, 0.8) rectangle (0.64, 0.6);

    \node (p13) at (-0.135, 1.3) {\includegraphics[width=.01\textwidth]{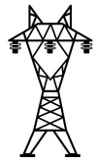}};
    \draw[rounded corners=1pt, line width=0.05pt] (-0.435, 1.53) rectangle (0.165, 1.07);
    \draw[rounded corners=1pt, line width=0.05pt, fill=white] (-0.085, 1.4) rectangle (0.515, 0.94);
    \node (p14) at (0.215, 1.17) {\includegraphics[width=.023\textwidth]{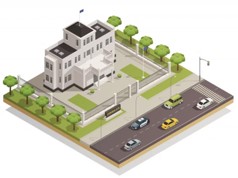}};
    \draw[rounded corners=1pt, line width=0.05pt] (-0.475, 1.57) rectangle (0.555, 0.9);
    \node[scale=0.13](ca) at (0.04,1.6){\textbf{Central authority} $L$};

    \node[scale=0.1](ci) at (0.04, 0.85){\textbf{Charging infrastructure}};
    \draw[-{Triangle[length=0.8pt, width=0.5pt]}, line width=0.05pt] (0.04, 0.6) -- (0.04, 0.4);

    \node[scale=0.13](pr) at (0.275,0.5){Prices $\pi\in\R^4$};
    \node[scale=0.13](rh) at (-0.625,0.43){\textbf{Ride-hailing market}};

\end{tikzpicture}}%
    \caption{Illustration of the problem setup with ride-hailing companies $\mc I=\{I_1, I_2, I_3\}$ operating in a region with charging stations $\mc M=\{M_1, M_2, M_3, M_4\}$. The central authority $L$ chooses the electricity price $\pi\in\mc P\subseteq\R^4$ so as to respect the discount budget $B_i$ of each company $i\in\mc I$.} 
\label{fig:market}
\end{figure}

\section{Numerical examples}\label{sec:5}
We consider two scenarios of a case study in the smart mobility domain previously introduced in~\cite{Hierarchical,ecc2023}. In particular, we analyze a market model depicted in Figure~\ref{fig:market}, where ride-hailing companies $\mc I=\{ I_1, I_2, I_3\}$ compete to meet demand requests that are distributed heterogeneously across the city of Shenzhen~\cite{nature}. The city is inherently partitioned into four Voronoi-based regions by the available charging infrastructure that consists of stations $\mc M=\{M_1, M_2, M_3, M_4\}$ and is controlled by the central authority $L$ through adjustable electricity prices $\pi\in[p_{\text{min}},p_{\text{max}}]^4$. At a particular point in time, we assume that each company $i\in\mc I$ wants to recharge its $N_i$ vehicles by distributing them among charging stations $\mc M$. Namely, we let the vector $x_i\in\mc X_i\subseteq \R^4$ denote the strategic decision of company $i$ that describes the fleet split among charging stations, i.e., $\norm{x_i}_1=N_i$ and $x_i^j\geq 0$ represents what fraction of the fleet is to be directed to a particular station $j\in\mc M$.    

The central authority, which can for example be the power-providing company or the government, may have an interest in balancing the demand on the power grid or it might aim to design pricing incentives to enhance coverage and encourage idle taxi drivers to avoid flocking to the more demand-attractive areas. We assume that the nominal prices of charging are encoded in $\pi_{\text{base}}\in\R^4$ and that the central authority is then interested in determining the optimal discount $\Delta\pi\in\R^4$, all while adhering to the total monetary discount budgets $B_i\in\R$ assigned to companies $\mc I$ based on external subsidies that they receive. Upon the announcement of the pricing vector $\pi\defineas\pi_{\text{base}}-\Delta\pi$, every company operator is interested in minimizing its operational cost
under the feasibility constraints imposed by the battery status of its vehicles. Similar to the objectives analyzed in~\cite{ecc2022, ecc2023, articleTH, 9541309}, the operator's cost is assumed in the form of a sum of three terms, i.e., $J_{i}(x_{i},\sigma(x_{-i}),\pi) = J^{1}_{i}(x_{i}, \sigma(x_{-i}))+J^{2}_{i}(x_{i})+J^{3}_{i}(x_{i}, \pi)$, where $J^{1}_{i}(x_{i}, \sigma(x_{-i}))$ denotes the expected queuing cost at different charging stations due to their limited capacities, $J^{2}_{i}(x_{i})$ denotes the negative expected revenue in the regions around charging stations, $J^{3}_{i}(x_{i}, \pi))$ denotes the charging cost and $\sigma(\cdot)$ is defined as in Definition~\ref{ex:1}. The resulting form is quadratic and given by $J_{i}(x_{i},\sigma(x_{-i}),\pi)=\frac{1}{2} x_{i}^T P_{i}x_{i}+x_{i}^TQ_{i}\sigma\left(x_{-i}\right)+r_{i}^{T}x_{i}+\pi^TS_ix_i$. On the other hand, we assume that the central authority chooses a desired vehicle distribution vector $\mc Z\in[0,1]^4$ satisfying $\norm{\mc Z}_1=1$ and plays the game with the ride-hailing companies in an attempt to minimize the cost $J_G (\sigma(x)) = \frac{1}{2}\|\sigma(x)-\textbf{1}^Tn\mc Z\|^{2}_{2}$, with $n=\text{col}((N_i)_{i\in\mc I})$ being the vector containing the number of vehicles per company that need to be recharged.

Concerning the constraint sets of ride-hailing companies, they encompass information about the
number of vehicles that can reach a certain station under a linear battery discharge model and given the current battery level after the rush-hour period simulation. It has been shown in~\cite{Hierarchical} that a specifically designed polytopic constraint allows for the consistent matching of each ride-hailing vehicle with precisely one charging station in an attempt to respect the allocation given by the split $x$. For every $i\in\mc I$, the matching constraints in accordance with~\cite{Hierarchical} are given by $\mc X_i^m\defineas\{x_i\in\R^{4}\mid A_ix_i=b_i \land G_ix_i\leq h_i\}$, for some properly chosen $A_i,b_i,G_i,h_i$. Apart from them, we also account for the limited discount budget $B_i$ through the constraint $\mc X_i^b(\pi)\defineas\{x_i\in\R^4 \mid (\pi_{\text{base}}-\pi)^TS_ix_i\leq B_i\}$. Hence, for any pricing strategy $\pi\in\mc P$ and for every $i\in\mc I$, the resulting constraint set is given by $\mc X_i(\pi)\defineas\mc X_i^m\cap\mc X_i^b(\pi)$. Generally speaking, $\mc X_i(\pi)$ is a polytopic constraint in $x_i$ but does not comply with the structure proposed in Assumption~\ref{ass:s1} of Section~\ref{sec:LQG}. Therefore, we test two scenarios:
\begin{enumerate}
    \item To illustrate the performance of the algorithm in a more general scenario, we shift away from the original setup in~\cite{ecc2023} and assume that the discount budgets are finite, i.e., $B_i<\infty$ for all $i\in\mc I$;
    \item To demonstrate the effects of the warm-start procedure, we set $B_i=\infty$ for all $i\in\mc I$, which yields $\mc X_i^b(\pi)=\R^4$ and gives rise to an identical problem setup as the one analyzed in~\cite{ecc2023}. 
\end{enumerate}
The number of vehicles per company that want to recharge is given by $n=[194,181,157]$ and $\mc Z$ is chosen to correspond to the total number of requests in each cell. For the analyzed case study, $\mc Z$ is such that $\textbf{1}^Tn\mc Z=\left[198, 103, 144, 87\right]$ and we set $p_{\text{min}}=0.0$ and $p_{\text{max}}=5.0$. For the extensive list of all remaining parameters in the simulation, we refer the reader to~\cite{Hierarchical}.

\subsection{Finite discount budgets}
\begin{figure}
    \centering
    \input{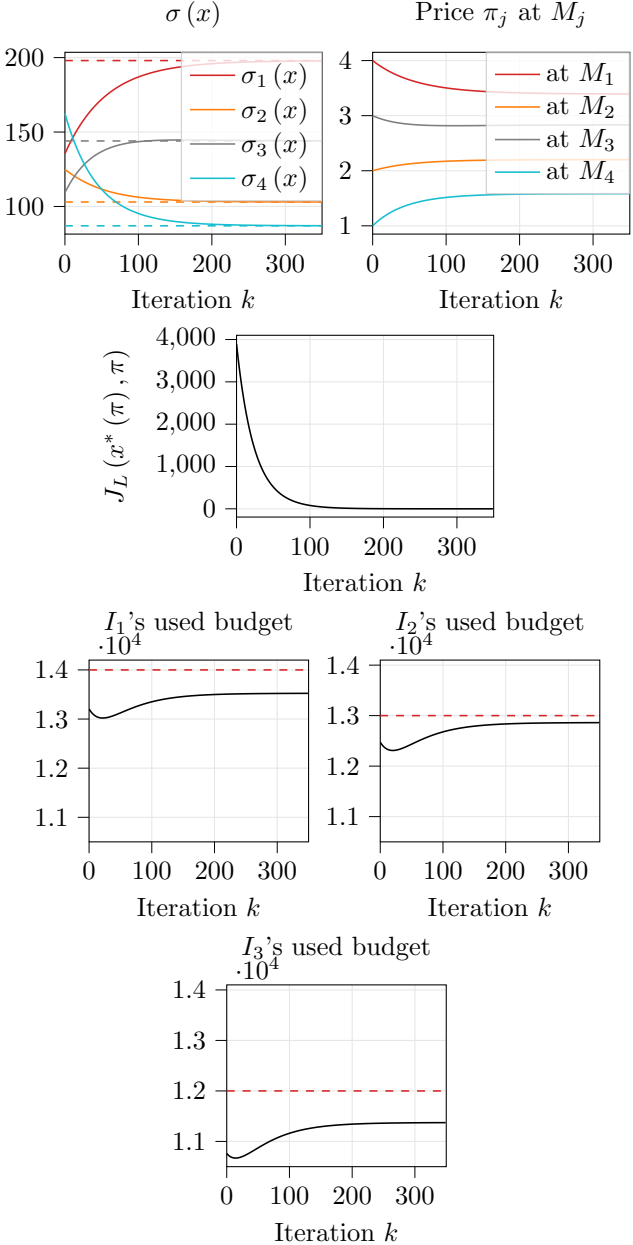}
    \caption{The plots show the evolution of the total vehicle accumulation at the charging stations $\sigma(x)$, the price of charging $\pi_j$ at the station $M_j$, the leader's objective $J_L(x^*(\pi),\pi)$, and the portion of the budgets used at each iteration.}
    \label{fig:accprice}
\end{figure}
In this case study, the finite discount budgets are given by vector $B=[14000,13000,12000]$ and the base price is given by $\pi_{\text{base}}=[5.0,3.0,5.0,3.0]$. Before each update step of the pricing policy, we perform $k_{\text{v-NE}}=5000$ of the Picard-Banach fixed point iteration procedure to compute the v-NE of the lower-level game~\cite{ecc2023} for the current value of the pricing vector. For the outer loop, we set the number of iterations to $k_{\text{l-SE}}=350$ and observe the average duration of one update step of approximately $\tau_{\text{avg}}\approx0.5 \text{ sec}$. For the given number of iterations and the pricing vector $\pi_{\text{init}}^1=[4.0,2.0,3.0,1.0]$ used to initialize the outer loop of the procedure, the system manages to achieve perfect matching with respect to the desired vehicle distribution and attains $J_L\left(x^*\left(\pi^*\right),\pi^*\right)=0.022$. This is further supported by plots in Figure~\ref{fig:accprice}. The three upper plots show the evolution of the attained vehicle accumulations at different charging stations, the evolution of the pricing vector $\pi$, and the corresponding value of the central authority's objective function. The lower three plots demonstrate that no discount budget constraint has been violated during the iterative procedure, i.e., the used discount budget for company $I_1$ is $B_1^{\text{used}}\approx13523$,  for company $I_2$ is $B_2^{\text{used}}\approx12860$, and for company $I_3$ is $B_3^{\text{used}}\approx11371$. A full overview of the relevant numerical values is presented in Table~\ref{tab:1}. It is important to note that the initial value $\pi_{\text{init}}$ has been obtained via sparse grid search as the setup does not comply with the structure of inequality constraints in Assumption~\ref{ass:s1} of Section~\ref{sec:LQG}. Since the complexity of the grid-search procedure grows exponentially in the size of $\pi$ and polynomially in the granularity of the grid, it is evident that this kind of heuristic is in general not suitable for larger problem sizes. However, for a broad class of bi-level games where the agent's constraints are given by~\eqref{eq:mcp} and~\eqref{eq:xipi}, we can deploy our iterative warm-up procedure. Therefore, in the following subsection, we will shift back our focus to the original problem setup of~\cite{ecc2023}.

\subsection{Infinite discount budgets}
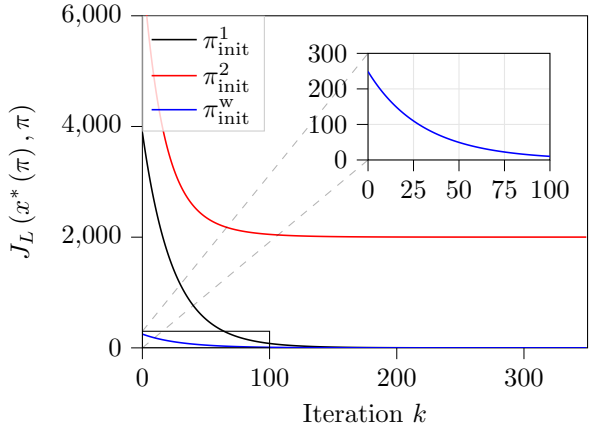
\begin{figure}
    \centering
    \input{figures/Warmup.tikz}
    \caption{Evolution of the central authority's objective for different initializations of the upper-level iterative loop.}%
    \label{fig:warm}
\end{figure}

\begin{table}
\begin{center}
 \renewcommand{\arraystretch}{1.2}
 \caption{Vehicle distribution and charging prices}\label{tab:1}
 \begin{tabular}{c|cc|cc}
 \multirow{2}{*}{Stations $\mc M$}  & \multicolumn{2}{c|}{Vehicle distribution} & \multicolumn{2}{c}{Charging prices}  \\ 
 &$\boldsymbol{1}^Tn\mc Z$ & $\sigma(x^*)$ &$\pi_{j}^*$ & $\Delta \pi_{j}^*$ \\
 \hline
 \hline
 $M_{1} \rule{0pt}{2.6ex}$ &  198 & 197.81 & 3.39  & 1.61   \\
 $M_{2}$                   &  103 & 103.04 & 2.20  & 0.80   \\
 $M_{3}$                   &  144 & 144.09 & 2.83  & 2.17   \\
 $M_{4}$                   &  87  &  87.06 & 1.58  & 1.42   \\ 
\hline 
\end{tabular}
\end{center}
\end{table}
When discount budgets are infinite for every ride-hailing company, starting the outer loop with the initial value $\pi_{\text{init}}^2=[3.0,3.0,3.0,3.0]$ yields that the value of the central authority's objective converges to $J_L\left(x^*\left(\pi^*\right),\pi^*\right)=2001$~\cite{ecc2023}. As previously discussed, $\pi_{\text{init}}^2$ already renders certain inequality constraints active which immediately creates a distinction between the original and the surrogate best-response optimization problems. Instead, we let the warm-start procedure with $\rho=1.0$ run for $k_{\text{w}}=500$ iterations to obtain the initial pricing vector $\pi_{\text{init}}^{\text{w}}=[4.8, 3.3, 3.9, 2.7]$. Starting the outer loop with $\pi_{\text{init}}^{\text{w}}$ induces an interior v-NE in the first iteration and the complete algorithm is later capable of recovering the perfect matching attained when starting from $\pi_{\text{init}}^1$. Since the theoretical optimal value for the central authority's objective is zero, the generation of two distinct pricing vectors from the initial states $\pi_{\text{init}}^1$ and $\pi_{\text{init}}^{\text{w}}$ indicates the general non-uniqueness of the solution in these bi-level games. In Figure~\ref{fig:warm}, we depict the evolution of the complete algorithm for different initial pricing vectors while Table~\ref{tab:2} lists all the relevant numerical values. It is interesting to note that the warm-start procedure provides a significantly smaller starting value of the central authority's objective compared to $\pi_{\text{init}}^1$ and $\pi_{\text{init}}^2$. However, from the perspective of ride-hailing company operators, starting from $\pi_{\text{init}}^1$ results in more favorable charging prices in terms of the pricing vector's magnitude and hence, the total charging costs.

\section{Conclusion}\label{sec:6}
\begin{table}
\begin{center}
 \renewcommand{\arraystretch}{1.2}
 \caption{Performance comparison for different initializations}\label{tab:2}
 \begin{tabular}{c|cccc|c}
 \multirow{2}{*}{Initial $\pi$}  & \multicolumn{4}{c|}{Resulting prices at $M_j$} & \multirow{2}{*}{Attained $J_L$}  \\ 
 & $\pi_1$ & $\pi_2$ & $\pi_3$ & $\pi_4$ & \\
 \hline
 \hline
 $\pi_{\text{init}}^1 \rule{0pt}{2.6ex}$ &  3.39 & 2.20 & 2.83  & 1.58 & $0.0222$  \\
 $\pi_{\text{init}}^2$                   &  3.57 & 2.38 & 3.01  & 3.00 & $2000.0$  \\
 $\pi_{\text{init}}^{\text{w}}$       &  4.58 & 3.36 & 3.99  & 2.75 & $0.0039$  \\ 
\hline 
\end{tabular}
\end{center}
\end{table}  
In this work, we formalized an iterative framework based on the Implicit Function Theorem for tackling the problem of computing the local Stackelberg equilibrium in bi-level games. Apart from generalizing the idea introduced in~\cite{ecc2023} to suit a broader class of games, we address the reported shortcomings of random initialization for a class of quadratic, aggregative games with polytopic constraints. In light of the overall decentralized nature of the approach, we formulate an internal v-NE feasibility test in the form of a linear program that can be efficiently solved by a distributed alternating direction method of multipliers. In addition to theoretical guarantees, we provide an experimental demonstration of the performance improvement in the previously analyzed case study in the smart mobility domain. 

In the future, it would be interesting to investigate if other distributed initialization methods could be designed that would be capable of tackling convex constraints beyond polyhedral form. Moreover, the proposed method allows for addressing various real-world problems in the domain of energy management, transportation, etc., hence making it interesting from the practical perspective as well.     

\appendix
\section{Appendix}
\begin{lemma}[Chapter 2.17 of~\cite{Bernstein2009}]\label{lem:3}
    Let $M_1\in\R^{p\times p}$, $M_2,M_3^T\in\R^{p\times q}$ and $M_4\in\R^{q\times q}$. If $M_1$ is nonsingular, then the inverse  
    $$M^{-1}\defineas\left[\begin{array}{cc}\overline{M}_1 & \overline{M}_2\\\overline{M}_3 & \overline{M}_4\end{array}\right]\:\:\text{of}\:\: M\defineas\left[\begin{array}{cc}M_1 & M_2\\M_3 & M_4\end{array}\right]$$
    exists if and only if Shur complement of $M_1$ in $M$, i.e., $\operatorname{Sh}\left(M_1\right)=M_4-M_3M_1^{-1}M_2$, is nonsingular. The blocks are given by $\overline{M}_1=M_1^{-1}+M_1^{-1}M_2\operatorname{Sh}(M_1)^{-1}M_3M_1^{-1}$, $\overline{M}_2=-M_1^{-1}M_2\operatorname{Sh}(M_1)^{-1}$, $\overline{M}_3=-\operatorname{Sh}(M_1)^{-1}M_3M_1^{-1}$ and $\overline{M}_4=\operatorname{Sh}(M_1)^{-1}$.
\end{lemma}

\bibliographystyle{plain} 
\bibliography{references.bib}  

\end{document}

%% file: figures/Warmup.tikz
\begin{tikzpicture}

\definecolor{color0}{rgb}{0.12156862745098,0.466666666666667,0.705882352941177}
\definecolor{color1}{rgb}{1,0.498039215686275,0.0549019607843137}
\definecolor{color2}{rgb}{0.0901960784313725,0.745098039215686,0.811764705882353}

\begin{axis}[
name=ax1,
legend cell align={left},
legend style={
  fill opacity=0.8,
  draw opacity=1,
  text opacity=1,
  at={(0.00,1.0)},
  anchor=north west,
  draw=white!80!black
},
tick align=outside,
tick pos=left,
width=7.5cm,
height=6cm,
xlabel={Iteration $k$},
xmin=0, xmax=350,
xtick style={color=black},
ylabel={$J_L\left(x^*\left(\pi\right),\pi\right)$},
ylabel style={yshift=+3pt},
ymin=0.0, ymax=6000,
ytick style={color=black}
]
\addplot [semithick, black]
table {%
0 3894.56810295853
1 3734.98998102338
2 3582.15251469823
3 3435.76358735916
4 3295.54402032471
5 3161.22698828496
6 3032.55746170797
7 2909.29167494514
8 2791.19661881979
9 2678.04955654322
10 2569.6375618586
11 2465.75707836709
12 2366.21349904127
13 2270.82076497913
14 2179.40098249787
15 2091.78405771031
16 2007.80734776775
17 1927.31532799273
18 1850.15927416186
19 1776.19695923508
20 1705.29236386035
21 1637.31540001572
22 1572.14164718036
23 1509.65210045541
24 1449.73293008289
25 1392.27525183681
26 1337.17490778599
27 1284.33225695078
28 1233.65197539955
29 1185.04286535113
30 1138.41767287016
31 1093.69291376192
32 1050.78870729108
33 1009.62861736672
34 970.139500852569
35 932.251362677329
36 895.897217434955
37 861.012957179351
38 827.537225131458
39 795.411295030091
40 764.578955869853
41 734.986401781902
42 706.582126824087
43 679.316824458059
44 653.143291501234
45 628.016336350967
46 603.892691287958
47 580.730928674584
48 558.491380872365
49 537.136063710779
50 516.628603347395
51 496.934166366664
52 478.019392971622
53 459.852333129231
54 442.402385536941
55 425.64023928365
56 409.537818084034
57 394.06822697106
58 379.205701336621
59 364.925558214698
60 351.204149707337
61 338.018818457102
62 325.347855074862
63 313.170457435557
64 301.466691758491
65 290.217455392682
66 279.40444123128
67 269.0101036825
68 259.017626127792
69 249.410889801093
70 240.174444026154
71 231.293477751475
72 222.753792325293
73 214.541775455735
74 206.644376303473
75 199.049081656885
76 191.743893141655
77 184.717305419195
78 177.958285330096
79 171.456251941006
80 165.201057454848
81 159.182968946552
82 153.392650887756
83 147.821148425788
84 142.459871383689
85 137.300578949755
86 132.335365025905
87 127.556644206496
88 122.957138359336
89 118.529863782984
90 114.268118914792
91 110.165472565648
92 106.215752658441
93 102.413035448142
94 98.7516352025341
95 95.2260943234069
96 91.831173889128
97 88.5618446000371
98 85.4132781092703
99 82.3808387223107
100 79.4600754489802
101 76.6467143927221
102 73.9366514626454
103 71.3259453940991
104 68.8108110644534
105 66.3876130915523
106 64.0528597022349
107 61.8031968595205
108 59.6354026371991
109 57.5463818310745
110 55.5331607968401
111 53.5928825046649
112 51.7228018011374
113 49.9202808698428
114 48.1827848818502
115 46.5078778280476
116 44.8932185254926
117 43.3365567903056
118 41.8357297700823
119 40.3886584288339
120 38.9933441781322
121 37.6478656479958
122 36.3503755916827
123 35.0990979188718
124 33.8923248512765
125 32.7284141959972
126 31.6057867313502
127 30.5229237003296
128 29.4783644074123
129 28.4707039140812
130 27.4985908289964
131 26.5607251887668
132 25.6558564255538
133 24.78278141786
134 23.9403426208264
135 23.1274262730076
136 22.3429606759564
137 21.5859145441318
138 20.8552954216429
139 20.1501481633168
140 19.4695534774219
141 18.8126265272367
142 18.1785155893522
143 17.5664007658852
144 16.9754927488611
145 16.4050316342255
146 15.8542857835346
147 15.3225507314128
148 14.809148136741
149 14.3134247759517
150 13.8347515765017
151 13.372522688911
152 12.9261545959162
153 12.49508525696
154 12.0787732868557
155 11.6766971668912
156 11.2883544873803
157 10.9132612200337
158 10.5509510192569
159 10.2009745509349
160 9.86289884772123
161 9.53630668964615
162 9.22079600916186
163 8.91597931947399
164 8.62148316527237
165 8.33694759503123
166 8.06202565383501
167 7.79638289604918
168 7.53969691698512
169 7.29165690275113
170 7.05196319758761
171 6.82032688806794
172 6.59646940324455
173 6.38012213042384
174 6.17102604565298
175 5.96893135857681
176 5.77359717079526
177 5.58479114753209
178 5.40228920180016
179 5.22587519067019
180 5.05534062321385
181 4.89048437959354
182 4.73111244080428
183 4.57703762883466
184 4.42807935652672
185 4.28406338718196
186 4.14482160298212
187 4.01019178250499
188 3.88001738634193
189 3.75414735105733
190 3.63243589070044
191 3.51474230604799
192 3.40093080065708
193 3.29087030419396
194 3.184434302093
195 3.08150067174938
196 2.98195152478002
197 2.88567305513789
198 2.79255539295627
199 2.70249246370076
200 2.61538185272002
201 2.53112467462779
202 2.44962544771988
203 2.37079197281855
204 2.29453521677715
205 2.22076920013933
206 2.14941088903288
207 2.08038009094525
208 2.01359935438813
209 1.94899387216719
210 1.88649138831533
211 1.82602210829646
212 1.76751861246157
213 1.71091577281913
214 1.65615067273393
215 1.60316252944176
216 1.55189261958003
217 1.50228420723579
218 1.45428247468226
219 1.40783445561829
220 1.36288897073609
221 1.31939656571922
222 1.27730945138319
223 1.23658144600267
224 1.19716791965038
225 1.15902574064967
226 1.12211322379881
227 1.08639008052705
228 1.05181737078965
229 1.01835745673088
230 0.985973957922397
231 0.954631708176748
232 0.924296714038064
233 0.894936114542361
234 0.866518142553105
235 0.839012087344599
236 0.812388258636929
237 0.786617951758672
238 0.761673414126562
239 0.737527812889311
240 0.714155203651899
241 0.691530500324006
242 0.66962944608531
243 0.6484285852639
244 0.62790523628064
245 0.608037465422967
246 0.588804061702831
247 0.570184512427659
248 0.55215897965536
249 0.534708277598838
250 0.517813850572566
251 0.501457751939597
252 0.4856226236152
253 0.470291676381748
254 0.455448670803889
255 0.441077898871299
256 0.427164166227158
257 0.413692775022355
258 0.400649507355411
259 0.388020609236264
260 0.375792775215814
261 0.363953133419272
262 0.352489231132495
263 0.341389020886709
264 0.330640847052564
265 0.320233432779787
266 0.310155867511639
267 0.300397594866809
268 0.290948400852358
269 0.281798402636923
270 0.272938037582207
271 0.264358052700118
272 0.25604949441913
273 0.248003698732646
274 0.240212281703862
275 0.232667130210757
276 0.225360393087612
277 0.218284472488449
278 0.211432015617902
279 0.204795906654908
280 0.198369259040192
281 0.192145407920179
282 0.186117902958358
283 0.18028050124849
284 0.174627160573436
285 0.16915203289318
286 0.163849457989272
287 0.158713957265718
288 0.153740227957314
289 0.14892313733435
290 0.144257717176515
291 0.139739158428711
292 0.135362806089688
293 0.13112415404612
294 0.127018840488745
295 0.123042642979271
296 0.119191474135732
297 0.115461377110478
298 0.111848521482898
299 0.108349199075747
300 0.104959820080694
301 0.101676909143862
302 0.0984971017896896
303 0.0954171407101967
304 0.0924338724216796
305 0.0895442438632017
306 0.0867452991260507
307 0.0840341764269397
308 0.0814081049684319
309 0.078864402101317
310 0.0764004704105901
311 0.0740137950415374
312 0.0717019410258217
313 0.0694625506948796
314 0.0672933412388375
315 0.0651921022799797
316 0.0631566935408046
317 0.0611850426284946
318 0.0592751428230258
319 0.0574250510471757
320 0.055632885738305
321 0.0538968249165919
322 0.0522151043114718
323 0.0505860154735274
324 0.049007904031896
325 0.0474791679152986
326 0.0459982557331386
327 0.04456366514205
328 0.043173941267014
329 0.0418276752061502
330 0.0405235025682487
331 0.039260102006665
332 0.0380361939460272
333 0.0368505391488725
334 0.035701937507838
335 0.0345892267687304
336 0.0335112812972511
337 0.0324670109985163
338 0.0314553601128864
339 0.030475306113658
340 0.0295258587721037
341 0.0286060590005945
342 0.0277149778994499
343 0.0268517158256145
344 0.0260154014904401
345 0.025205190955603
346 0.0244202668982325
347 0.0236598376686743
348 0.022923136530153
349 0.0222094208438648
};
\addlegendentry{$\pi_{\text{init}}^1$}
\addplot [semithick, red]
table {%
0 6677.83929746328
1 6512.98016016162
2 6351.54002783648
3 6104.76628723157
4 5872.85128834606
5 5654.86916885333
6 5449.952823225
7 5257.29013875311
8 5076.12047383381
9 4905.73136288334
10 4745.4554332662
11 4594.66752056034
12 4452.78196936645
13 4319.25010769581
14 4193.55788374243
15 4075.22365456948
16 3963.7961169154
17 3858.85237095711
18 3759.99610846036
19 3666.85591729965
20 3579.08369484827
21 3496.35316322301
22 3418.35847982058
23 3344.81293700736
24 3275.44774521896
25 3210.01089409814
26 3148.26608664511
27 3089.99174167945
28 3034.98006021561
29 2983.03615163804
30 2933.97721582735
31 2887.63177763714
32 2843.83897035338
33 2802.4478649856
34 2763.31684244215
35 2726.31300583199
36 2691.31163031353
37 2658.19564807666
38 2626.85516620078
39 2597.18701527604
40 2569.09432681202
41 2542.48613758519
42 2517.277019195
43 2493.3867312113
44 2470.73989639845
45 2449.26569660017
46 2428.89758795987
47 2409.57303423662
48 2391.23325705685
49 2373.8230020165
50 2357.29031961822
51 2341.58636009345
52 2326.66518122075
53 2312.48356830805
54 2299.00086556152
55 2286.17881811198
56 2273.98142401808
57 2262.37479560853
58 2251.32702956692
59 2240.8080852008
60 2230.78967037272
61 2221.24513460468
62 2212.14936889816
63 2203.47871184209
64 2195.21086160818
65 2187.32479345852
66 2179.80068241489
67 2172.61983076164
68 2165.7646000745
69 2159.21834748807
70 2152.96536593304
71 2146.9908280908
72 2141.28073383019
73 2135.82186090561
74 2130.60171870969
75 2125.60850488789
76 2120.83106463317
77 2116.25885249241
78 2111.88189652507
79 2107.69076466651
80 2103.67653315637
81 2099.83075690241
82 2096.14544165782
83 2092.6130178981
84 2089.22631629062
85 2085.97854465729
86 2082.86326633632
87 2079.87437985581
88 2077.00609983691
89 2074.25293904998
90 2071.60969155149
91 2069.0714168346
92 2066.6334249301
93 2064.29126239878
94 2062.0406991597
95 2059.87771610286
96 2057.79849343716
97 2055.79939972886
98 2053.87698158715
99 2052.02795395776
100 2050.24919098641
101 2048.53771741759
102 2046.89070049556
103 2045.30544233712
104 2043.77937274675
105 2042.31004244777
106 2040.8951167037
107 2039.53236930633
108 2038.21967690806
109 2036.955013678
110 2035.73644626176
111 2034.56212902713
112 2033.43029957802
113 2032.3392745209
114 2031.2874454683
115 2030.27327526554
116 2029.29529442691
117 2028.35209776952
118 2027.44234123237
119 2026.56473887018
120 2025.71806001145
121 2024.90112657128
122 2024.11281050952
123 2023.35203142607
124 2022.61775428542
125 2021.90898726231
126 2021.22477970227
127 2020.56422018994
128 2019.92643471886
129 2019.3105849574
130 2018.71586660491
131 2018.14150783296
132 2017.5867678072
133 2017.0509352848
134 2016.53332728354
135 2016.03328781851
136 2015.55018670236
137 2015.08341840595
138 2014.63240097574
139 2014.19657500494
140 2013.77540265542
141 2013.36836672757
142 2012.97496977561
143 2012.59473326569
144 2012.22719677463
145 2011.87191722712
146 2011.52846816911
147 2011.19643907575
148 2010.87543469181
149 2010.5650744031
150 2010.26499163691
151 2009.97483329039
152 2009.69425918505
153 2009.42294154628
154 2009.16056450634
155 2008.90682363004
156 2008.66142546141
157 2008.4240870908
158 2008.19453574097
159 2007.97250837168
160 2007.75775130124
161 2007.55001984492
162 2007.34907796861
163 2007.15469795771
164 2006.96666010001
165 2006.78475238204
166 2006.60877019831
167 2006.43851607278
168 2006.27379939183
169 2006.11443614853
170 2005.96024869735
171 2005.81106551883
172 2005.66672099415
173 2005.52705518858
174 2005.39191364383
175 2005.26114717874
176 2005.13461169778
177 2005.01216800741
178 2004.89368163951
179 2004.77902268181
180 2004.66806561503
181 2004.5606891562
182 2004.45677610827
183 2004.35621321531
184 2004.25889102345
185 2004.16470374697
186 2004.07354913963
187 2003.98532837079
188 2003.89994590616
189 2003.81730939315
190 2003.73732955039
191 2003.65992006145
192 2003.5849974724
193 2003.51248109323
194 2003.4422929029
195 2003.37435745775
196 2003.30860180344
197 2003.2449553899
198 2003.18334998949
199 2003.12371961807
200 2003.06600045889
201 2003.01013078922
202 2002.9560509096
203 2002.90370307564
204 2002.85303143215
205 2002.80398194976
206 2002.75650236357
207 2002.71054211407
208 2002.66605229022
209 2002.62298557431
210 2002.5812961889
211 2002.54093984557
212 2002.50187369539
213 2002.46405628118
214 2002.42744749139
215 2002.39200851554
216 2002.35770180119
217 2002.32449101245
218 2002.29234098985
219 2002.26121771155
220 2002.23108825597
221 2002.20192076555
222 2002.17368441191
223 2002.14634936202
224 2002.11988674563
225 2002.09426862374
226 2002.06946795814
227 2002.04545858195
228 2002.02221517125
229 2001.99971321753
230 2001.9779290011
231 2001.95683956545
232 2001.93642269241
233 2001.91665687807
234 2001.89752130968
235 2001.87899584313
236 2001.86106098125
237 2001.84369785294
238 2001.82688819271
239 2001.81061432122
240 2001.79485912621
241 2001.77960604423
242 2001.76483904279
243 2001.75054260335
244 2001.7367017046
245 2001.72330180642
246 2001.71032883447
247 2001.69776916499
248 2001.68560961045
249 2001.67383740534
250 2001.6624401927
251 2001.65140601093
252 2001.64072328099
253 2001.63038079425
254 2001.62036770042
255 2001.61067349612
256 2001.60128801373
257 2001.59220141055
258 2001.5834041584
259 2001.57488703362
260 2001.56664110709
261 2001.55865773494
262 2001.55092854936
263 2001.54344544972
264 2001.53620059407
265 2001.52918639076
266 2001.52239549049
267 2001.51582077853
268 2001.50945536717
269 2001.50329258856
270 2001.49732598753
271 2001.49154931484
272 2001.48595652065
273 2001.48054174806
274 2001.475299327
275 2001.47022376819
276 2001.46530975742
277 2001.46055214987
278 2001.45594596487
279 2001.45148638044
280 2001.44716872835
281 2001.4429884892
282 2001.43894128758
283 2001.43502288758
284 2001.43122918828
285 2001.42755621945
286 2001.42400013738
287 2001.42055722081
288 2001.41722386706
289 2001.41399658825
290 2001.41087200761
291 2001.40784685592
292 2001.40491796815
293 2001.40208228009
294 2001.39933682504
295 2001.3966787309
296 2001.39410521703
297 2001.3916135913
298 2001.38920124735
299 2001.38686566183
300 2001.3846043917
301 2001.3824150718
302 2001.38029541221
303 2001.37824319594
304 2001.3762562766
305 2001.37433257616
306 2001.37247008269
307 2001.37066684839
308 2001.36892098741
309 2001.367230674
310 2001.36559414046
311 2001.36400967542
312 2001.36247562197
313 2001.36099037593
314 2001.35955238422
315 2001.35816014315
316 2001.35681219695
317 2001.35550713613
318 2001.35424359608
319 2001.35302025566
320 2001.35183583573
321 2001.35068909789
322 2001.34957884314
323 2001.34850391065
324 2001.34746317655
325 2001.3464555527
326 2001.34547998561
327 2001.34453545534
328 2001.34362097434
329 2001.34273558656
330 2001.34187836635
331 2001.3410484175
332 2001.34024487235
333 2001.33946689081
334 2001.33871365957
335 2001.33798439115
336 2001.33727832318
337 2001.3365947175
338 2001.33593285947
339 2001.33529205722
340 2001.33467164081
341 2001.33407096168
342 2001.33348939189
343 2001.33292632348
344 2001.33238116784
345 2001.33185335509
346 2001.33134233345
347 2001.33084756877
348 2001.33036854379
349 2001.32990475781
};
\addlegendentry{$\pi_{\text{init}}^2$}
\addplot [semithick, blue]
table {%
0 248.822089970625
1 240.617838526727
2 232.706152099097
3 225.074902010601
4 217.712584254685
5 210.608279730011
6 203.751617389982
7 197.132740077705
8 190.7422728356
9 184.571293496021
10 178.611305374459
11 172.854211902199
12 167.292293047634
13 161.918183388192
14 156.724851705669
15 151.70558198799
16 146.85395573005
17 142.163835434676
18 137.629349222658
19 133.244876468256
20 129.005034383161
21 124.904665477941
22 120.938825835699
23 117.102774137962
24 113.391961387322
25 109.802021275915
26 106.328761152778
27 102.968153546703
28 99.7163282048459
29 96.5695646101231
30 93.5242849435781
31 90.5770474603196
32 87.7245402502595
33 84.9635753569455
34 82.2910832296511
35 79.7041074864683
36 77.1997999667583
37 74.7754160541263
38 72.428310251511
39 70.155931992078
40 67.9558216703517
41 65.8256068793999
42 63.7629988409681
43 61.76578901627
44 59.8318458860776
45 57.9591118897588
46 56.1456005132786
47 54.3893935174492
48 52.6886382976809
49 51.0415453676105
50 49.4463859594434
51 47.9014897339875
52 46.4052425943082
53 44.9560845970773
54 43.552507956203
55 42.1930551335499
56 40.8763170120401
57 39.6009311468442
58 38.3655800902925
59 37.1689897868164
60 36.0099280341856
61 34.887203007569
62 33.7996618437028
63 32.7461892814135
64 31.7257063564539
65 30.7371691475055
66 29.7795675712114
67 28.8519242237344
68 27.9532932666771
69 27.0827593553877
70 26.2394366077388
71 25.4224676113336
72 24.6310224677582
73 23.8642978719399
74 23.1215162252847
75 22.4019247809956
76 21.7047948203508
77 21.0294208585183
78 20.3751198786376
79 19.7412305932121
80 19.1271127314576
81 18.5321463516666
82 17.9557311776225
83 17.3972859580026
84 16.8562478479944
85 16.3320718120885
86 15.8242300474849
87 15.3322114269504
88 14.8555209608749
89 14.3936792772984
90 13.9462221195681
91 13.5126998609521
92 13.092677035329
93 12.6857318837283
94 12.2914559158016
95 11.9094534859869
96 11.5393413835736
97 11.1807484363308
98 10.8333151272745
99 10.4966932238385
100 10.1705454193543
101 9.85454498607214
102 9.54837543975009
103 9.25173021488081
104 8.9643123506321
105 8.6858341869156
106 8.41601707035807
107 8.15459106959315
108 7.90129469991734
109 7.65587465678618
110 7.41808555773605
111 7.18768969278972
112 6.96445678269447
113 6.74816374503644
114 6.53859446764181
115 6.33553958955235
116 6.13879628861832
117 5.94816807601819
118 5.76346459747583
119 5.58450144048402
120 5.4110999479235
121 5.24308703748829
122 5.08029502677164
123 4.92256146404543
124 4.76972896436928
125 4.6216450507236
126 4.4781620003705
127 4.33913669596222
128 4.20443048137895
129 4.07390902213228
130 3.94744217016341
131 3.82490383279583
132 3.70617184611183
133 3.59112785190882
134 3.47965717889747
135 3.37164872742505
136 3.26699485788777
137 3.16559128261724
138 3.06733696125957
139 2.97213399932298
140 2.87988754997423
141 2.79050571888365
142 2.70389947213334
143 2.61998254689388
144 2.53867136500412
145 2.4598849492213
146 2.38354484205411
147 2.30957502722231
148 2.23790185350663
149 2.16845396105418
150 2.10116220985947
151 2.03595961064639
152 1.97278125780576
153 1.91156426447924
154 1.85224769958586
155 1.79477252697325
156 1.73908154638411
157 1.68511933612172
158 1.63283219785808
159 1.58216810283193
160 1.53307663983287
161 1.48550896496454
162 1.43941775267376
163 1.39475714864966
164 1.3514827238796
165 1.30955143037136
166 1.26892155810856
167 1.2295526934613
168 1.19140567877548
169 1.15444257336276
170 1.1186266155928
171 1.08392218615336
172 1.05029477265271
173 1.01771093501156
174 0.986138272241078
175 0.955545390006591
176 0.925901869431982
177 0.897178236671607
178 0.869345933551813
179 0.842377289067372
180 0.816245491856534
181 0.790924563360022
182 0.76638933205686
183 0.742615408296842
184 0.719579159933346
185 0.697257688887476
186 0.675628808301553
187 0.654671020351088
188 0.634363494900754
189 0.614686048735166
190 0.595619125422672
191 0.577143775855802
192 0.559241639399261
193 0.541894925554516
194 0.525086396224651
195 0.50879934868135
196 0.493017598731967
197 0.477725464770629
198 0.462907752033061
199 0.448549737597205
200 0.43463715560938
201 0.421156183150742
202 0.408093426467531
203 0.395435907612409
204 0.383171051526006
205 0.371286673569557
206 0.359770967344957
207 0.348612492980465
208 0.337800165674707
209 0.327323244800937
210 0.31717132306585
211 0.30733431626868
212 0.297802453234908
213 0.28856626601555
214 0.279616580628499
215 0.270944507770764
216 0.262541434101877
217 0.25439901356367
218 0.246509159162088
219 0.238864034916332
220 0.231456048004475
221 0.224277841240109
222 0.217322285840055
223 0.210582474188413
224 0.204051713055378
225 0.19772351694337
226 0.191591601615073
227 0.185649877861579
228 0.179892445379664
229 0.174313587049255
230 0.168907763083553
231 0.163669605652103
232 0.158593913420191
233 0.153675646561169
234 0.148909921550512
235 0.144292006410979
236 0.139817315997789
237 0.135481407458428
238 0.131279975732468
239 0.127208849426097
240 0.123263986493839
241 0.119441470349557
242 0.115737505879224
243 0.112148415766569
244 0.108670636789611
245 0.105300716208149
246 0.102035308489576
247 0.0988711718455306
248 0.0958051650704874
249 0.0928342444458394
250 0.08995546062215
251 0.0871659557778912
252 0.0844629607745446
253 0.0818437923589954
254 0.0793058505223598
255 0.0768466159097443
256 0.0744636473282299
257 0.0721545792948746
258 0.0699171197593387
259 0.0677490477100946
260 0.065648211104417
261 0.06361252458737
262 0.0616399675345747
263 0.0597285819858371
264 0.0578764707206574
265 0.0560817953955848
266 0.0543427746488305
267 0.0526576824013318
268 0.0510248461068841
269 0.0494426450968604
270 0.0479095089904149
271 0.0464239160974103
272 0.0449843919668638
273 0.0435895077971509
274 0.0422378792136442
275 0.0409281646898307
276 0.0396590643795207
277 0.0384293186834839
278 0.0372377071653318
279 0.0360830471290683
280 0.0349641926513868
281 0.033880033337482
282 0.0328294931823621
283 0.0318115296067845
284 0.0308251323513105
285 0.0298693224722228
286 0.0289431513956515
287 0.0280457000371825
288 0.0271760776959127
289 0.0263334213668713
290 0.0255168948278879
291 0.0247256877519249
292 0.0239590149540163
293 0.0232161156091024
294 0.0224962524807779
295 0.0217987111900584
296 0.0211227995569061
297 0.0204678467925987
298 0.0198332030013262
299 0.0192182383434556
300 0.0186223426208016
301 0.0180449244471674
302 0.0174854108772706
303 0.0169432466827857
304 0.0164178938575787
305 0.0159088310902007
306 0.0154155532291043
307 0.014937570780603
308 0.0144744094541238
309 0.0140256096638041
310 0.0135907260446402
311 0.0131693270886899
312 0.0127609946430312
313 0.0123653235095844
314 0.0119819210995047
315 0.0116104069529683
316 0.0112504124554107
317 0.0109015803609509
318 0.0105635645741131
319 0.0102360296696133
320 0.00991865065589081
321 0.00961111261858605
322 0.00931311040767469
323 0.00902434832823928
324 0.00874453985670698
325 0.00847340735708713
326 0.00821068181176088
327 0.00795610253771883
328 0.00770941692826455
329 0.00747038020926993
330 0.00723875522453454
331 0.00701431213019532
332 0.00679682824920746
333 0.00658608778394409
334 0.00638188166340115
335 0.00618400722305523
336 0.0059922681612079
337 0.0058064742370334
338 0.00562644108140375
339 0.00545199011321529
340 0.00528294821560849
341 0.00511914765593247
342 0.00496042593658785
343 0.00480662558038603
344 0.004657593977754
345 0.00451318324485328
346 0.00437325014718226
347 0.00423765580126201
348 0.00410626570737804
349 0.00397894949492184
};
\addlegendentry{$\pi_{\text{init}}^{\text{w}}$}

  \coordinate (c1) at (axis cs:0.0, 0.0);
  \coordinate (c2) at (axis cs:0.0, 300);
  \draw (c1) rectangle (axis cs:100,300);

\end{axis}

 \begin{axis}[
 xtick={0.0, 25, 50, 75, 100},
xticklabels={%
$0$,
$25$,
$50$,
$75$,
$100$,
},
   name=ax2,
tick align=outside,
tick pos=left,
width=4cm,
height=3cm,
x grid style={white!90!black},
xmajorgrids,
xmin=0, xmax=100,
ymin=0.0, ymax=300,
xtick style={color=black},
y grid style={white!90!black},
ymajorgrids,
ytick style={color=black}, 
   at={($(ax1.south west)+(3cm,2.5cm)$)},
 ]

\addplot [semithick, blue]
table {%
0 248.822089970625
1 240.617838526727
2 232.706152099097
3 225.074902010601
4 217.712584254685
5 210.608279730011
6 203.751617389982
7 197.132740077705
8 190.7422728356
9 184.571293496021
10 178.611305374459
11 172.854211902199
12 167.292293047634
13 161.918183388192
14 156.724851705669
15 151.70558198799
16 146.85395573005
17 142.163835434676
18 137.629349222658
19 133.244876468256
20 129.005034383161
21 124.904665477941
22 120.938825835699
23 117.102774137962
24 113.391961387322
25 109.802021275915
26 106.328761152778
27 102.968153546703
28 99.7163282048459
29 96.5695646101231
30 93.5242849435781
31 90.5770474603196
32 87.7245402502595
33 84.9635753569455
34 82.2910832296511
35 79.7041074864683
36 77.1997999667583
37 74.7754160541263
38 72.428310251511
39 70.155931992078
40 67.9558216703517
41 65.8256068793999
42 63.7629988409681
43 61.76578901627
44 59.8318458860776
45 57.9591118897588
46 56.1456005132786
47 54.3893935174492
48 52.6886382976809
49 51.0415453676105
50 49.4463859594434
51 47.9014897339875
52 46.4052425943082
53 44.9560845970773
54 43.552507956203
55 42.1930551335499
56 40.8763170120401
57 39.6009311468442
58 38.3655800902925
59 37.1689897868164
60 36.0099280341856
61 34.887203007569
62 33.7996618437028
63 32.7461892814135
64 31.7257063564539
65 30.7371691475055
66 29.7795675712114
67 28.8519242237344
68 27.9532932666771
69 27.0827593553877
70 26.2394366077388
71 25.4224676113336
72 24.6310224677582
73 23.8642978719399
74 23.1215162252847
75 22.4019247809956
76 21.7047948203508
77 21.0294208585183
78 20.3751198786376
79 19.7412305932121
80 19.1271127314576
81 18.5321463516666
82 17.9557311776225
83 17.3972859580026
84 16.8562478479944
85 16.3320718120885
86 15.8242300474849
87 15.3322114269504
88 14.8555209608749
89 14.3936792772984
90 13.9462221195681
91 13.5126998609521
92 13.092677035329
93 12.6857318837283
94 12.2914559158016
95 11.9094534859869
96 11.5393413835736
97 11.1807484363308
98 10.8333151272745
99 10.4966932238385
100 10.1705454193543
};%

\end{axis}

\draw [dashed, opacity=0.3] (c1) -- (ax2.south west);
\draw [dashed, opacity=0.3] (c2) -- (ax2.north west);
\end{tikzpicture}